\documentclass[reqno,12pt]{amsart}
\usepackage{amssymb, latexsym, euscript}
\newtheorem{theorem}{Theorem}[section]
\newtheorem{proposition}[theorem]{Proposition}

\newtheorem{lemma}[theorem]{Lemma}
\newtheorem{definition}[theorem]{Definition}
\newtheorem{corollary}[theorem]{Corollary}

\theoremstyle{definition}

\newtheorem{remark}[theorem]{Remark}

\numberwithin{equation}{section}

\begin{document}
\title[Lax-Phillips scattering scheme for $\mathcal{PT}$-symmetric operators]{On elements of the Lax-Phillips scattering scheme for $\mathcal{PT}$-symmetric operators}
\author[S.~Albeverio]{Sergio~Albeverio}
\author[S.~Kuzhel]{Sergii~Kuzhel}

\address{Universit\"{a}t Bonn, Institut f\"{u}r Angewandte Mathematik,
Endenicher Allee 60, D-53115 Bonn, Germany; SFB
611, Bonn University; IZKS, Bonn University; BiBoS (Bielefeld-Bonn).}
\email{albeverio@uni-bonn.de}

\address{AGH University of Science and Technology \\ Department of Applied Mathematics \\
30-059 Krakow, Poland} \email{kuzhel@mat.agh.edu.pl}


\maketitle
\begin{abstract}
Generalized $\mathcal{PT}$-symmetric operators acting an a Hilbert space $\mathfrak{H}$ are defined and investigated.
The case of $\mathcal{PT}$-symmetric extensions of a symmetric operator $S$ is investigated in detail.
The possible application of the Lax-Phillips scattering methods to the investigation of $\mathcal{PT}$-symmetric
operators is illustrated by considering the case of $0$-perturbed operators.
\end{abstract}

\section{Introduction}
The employing of non-self-adjoint operators for the description of experimentally observable data
goes back to the early days of quantum mechanics. Nowadays, the steady interest in this subject grew considerably after it has been discovered numerically \cite{B1} and rigorously proved  \cite{D2} that the spectrum of the Hamiltonian
\begin{equation}\label{e1b}
    H=-\frac{d^2}{dx^2} + x^2(ix)^\epsilon, \qquad  0\leq\epsilon<2
\end{equation}
 is real and positive. It was conjectured \cite{B1} that the reality of the spectrum of $H$ is a consequence of its
$\mathcal{P}\mathcal{T}$-symmetry:
$${\mathcal P}{\mathcal
T}H=H{\mathcal P}{\mathcal T},
$$ where the space reflection
(parity) operator $\mathcal{P}$ and the complex conjugation operator
$\mathcal{T}$ are defined as follows:
$({\mathcal P}f)(x)=f(-x)$ and $({\mathcal T}f)(x)=\overline{f(x)}.$
 This gave rise to a consistent complex extension of conventional
quantum mechanics into $\mathcal{PT}$ quantum mechanics (PTQM),
see e.g. the review papers \cite{BE,MO} and the references therein.

In general, the Hamiltonians of PTQM are not self-adjoint with respect to the initial Hilbert space's inner product
but they possess a certain `more physical property' of symmetry, which does not depend on the choice of inner product
(like the above $\mathcal{PT}$-symmetry property).

Typically, $\mathcal{P}\mathcal{T}$-symmetric Hamiltonians
can be interpreted as self-adjoint ones for a suitable choice of Krein space
(for the definition of Krein space see Sect. 2).
For instance, the $\mathcal{PT}$-symmetric operator $H$ in (\ref{e1b})
turns out to be  self-adjoint with respect to the indefinite metric
$$
[f,g]_{\mathcal{P}}:=(\mathcal{P}f,g)=\int_{-\infty}^{\infty}f(-x)\overline{g(x)}dx, \qquad f,g\in{L_2(\mathbb{R})}.
$$
The space $L_2(\mathbb{R})$ with the indefinite metric $[\cdot,\cdot]_{\mathcal{P}}$ is the Krein space $(L_2(\mathbb{R}), [\cdot,\cdot]_{\mathcal{P}})$. Hence, $H$ is self-adjoint in the Krein space  $(L_2(\mathbb{R}), [\cdot,\cdot]_{\mathcal{P}})$.

However the self-adjointness of $H$ in a Krein space cannot
be satisfactory because it does not guarantee the unitarity of
the dynamics generated by $H$.  To do so one must demonstrate that $H$ is self-adjoint on a \emph{Hilbert space} (not on a Krein space!).
This problem can be overcome for an operator $H$, which is self-adjoint with respect to an indefinite metric $[\cdot,\cdot]$, by finding a new symmetry represented by a linear operator $\mathcal{C}$ and such that the semilinear form $(\cdot,\cdot)_{\mathcal{C}}:=[\mathcal{C}\cdot,\cdot]$ is a
(positively defined) inner product and $H$ turns out to be self-adjoint with respect to $(\cdot,\cdot)_{\mathcal{C}}$ \cite{BE, BBJ1}.

The description of a symmetry $\mathcal{C}$ for a given $\mathcal{PT}$-symmetric Hamiltonian $H$ is one of the key points in PTQM.
Because of the complexity of the problem (since $\mathcal{C}$ depends on the choice of $H$), it is not
surprising that the majority of the available formulae are still approximative, usually
expressed as leading terms of perturbation series \cite{BE,BMY}.

In the present paper we are going to investigate the problems mentioned above for
$\mathcal{PT}$-symmetric operators in some abstract setting. This is a natural problem because
various Hamiltonians $H$ of PTQM have the property of $\mathcal{PT}$-symmetry realized by different operators
$\mathcal{P}$ and $\mathcal{T}$.

The concept of $\mathcal{PT}$-symmetry can be easily formulated for the general case of a linear densely defined operator $H$ in an abstract Hilbert space $\mathfrak{H}$ with the use of an unitary involution $\mathcal{P}$ and a conjugation $\mathcal{T}$ operator both acting in
$\mathfrak{H}$; see, Subsection 2.1 and Definition \ref{ddd1}. Beginning with this definition we study $\mathcal{PT}$-symmetric extensions
of a symmetric operator $S$ acting in a Hilbert space $\mathfrak{H}$ assuming additionally that $S$ commutes with the elements of the
Clifford algebra $\mathcal{C}l_2(\mathcal{P}, \mathcal{R})$. Such kind of restrictions are motivated by the $\mathcal{PT}$-symmetric
Hamiltonians appearing due to non-self-adjoint boundary conditions for Schr\"{o}dinger operators with singular potentials \cite{AlKur,AK,AlKuz,AK,TFC}.

The Clifford algebras technique is relevant for $\mathcal{PT}$-symmetric studies \cite{GK} and it allows one to
interpret $\mathcal{PT}$-symmetric extensions of $S$ as self-adjoint operators in Krein spaces
$(\mathfrak{H}, [\cdot,\cdot]_{\mathcal{P}_\xi})$ with involutions ${\mathcal{P}_\xi}$ constructed in terms of $\mathcal{C}l_2(\mathcal{P}, \mathcal{R})$; see Theorem \ref{es16}.

Our studies show that the `physical' condition of $\mathcal{PT}$-symmetry leads, in many cases, to a simplification of the
mathematical results. In particular, the condition of $\mathcal{PT}$-symmetry imposed on operators $\mathcal{C}$ from $\mathcal{C}l_2(\mathcal{P}, \mathcal{R})$ leads to the very simple presentation
$\mathcal{C}=e^{\chi{i}\mathcal{R}\mathcal{P}_\xi}\mathcal{P}_\xi$ given by Lemma \ref{ll2}. This allows one to establish
a clear relationship between $\mathcal{PT}$-symmetric extensions $H$ of $S$ and the corresponding $\mathcal{C}$-symmetries,
see Theorem \ref{esse8}.

In the second part of the present paper we apply the Lax-Phillips scattering method \cite{LF} to the investigation of $\mathcal{PT}$-symmetric operators. Our approach is based on an operator-theoretical interpretation of the Lax-Phillips scattering scheme developed in \cite{KU1,AlAn}.
We present explicit formulas for an analytical continuation of the scattering matrix and consider in detail the case of $0$-perturbed
$\mathcal{PT}$-symmetric operators, see Section 4.

%Theorem \ref{esse3} deals with the general case and, after some technical work, it may give a recipe for the construction of analytical continuation %of the scattering matrix for Schr\"{o}dinger operators with
%finite supported $\mathcal{PT}$-symmetric potentials (like \cite{Jones}). We are aiming at studying this point in a forthcoming publication.

Throughout the paper, $\mathcal{D}(A)$ and $\mathcal{R}(A)$ denote the domain and the range of a linear operator $A$, respectively.  $A\upharpoonright_{\mathcal{D}}$ means the restriction of $A$ onto a set $\mathcal{D}$.

\section{Abstract $\mathcal{PT}$-symmetric operators}
\subsection{Preliminaries.}
Let $\mathfrak{H}$ be a complex Hilbert space with inner product
$(\cdot,\cdot)$ linear in the first argument. A linear operator $\mathcal{P}$ defined on $\mathfrak{H}$ is called \emph{unitary involution} if
\begin{equation}\label{bonn1}
(i) \quad \mathcal{P}^2=I, \qquad (ii) \quad (\mathcal{P}f,\mathcal{P}g)=(f,g)
\end{equation}
for all $f,g\in\mathfrak{H}$. It follows from (\ref{bonn1}) that $\mathcal{P}$ is  simultaneously self-adjoint and unitary.

Let $\mathcal{J}$ be a unitary involution in $\mathfrak{H}$.
The Hilbert space $\mathfrak{H}$ endowed with the indefinite metric
$$
[f,g]_{\mathcal{J}}:=(\mathcal{J}f,g)
$$
is a called a Krein space \cite{AZ,BO} and it will be denoted by $(\mathfrak{H}, [\cdot,\cdot]_{\mathcal{J}})$.

A modification of condition (ii) in (\ref{bonn1}) leads to the definition of the conjugation operator.
An operator $\mathcal{T}$ defined on $\mathfrak{H}$ is called \emph{conjugation} if
\begin{equation}\label{bonn2}
(i) \quad \mathcal{T}^2=I, \qquad (ii) \quad (\mathcal{T}f,\mathcal{T}g)=(g,f)
\end{equation}
for all $f,g\in\mathfrak{H}$. The conjugation operator is bounded in $\mathfrak{H}$ but, in contrast to the case of an involution, it is \emph{anti-linear} in the sense that \cite{AkGl}
\begin{equation}\label{bonn2b}
 \mathcal{T}(\alpha{f}+\beta{g})=\overline{\alpha}\mathcal{T}{f}+\overline{\beta}\mathcal{T}{g}, \qquad \alpha, \beta\in\mathbb{C}, \quad f,g\in\mathfrak{H}.
\end{equation}

Let us fix a unitary involution $\mathcal{P}$ and a conjugation $\mathcal{T}$ in $\mathfrak{H}$ \emph{assuming that
$\mathcal{P}$ and $\mathcal{T}$ commute}, that is, $\mathcal{P}\mathcal{T}=\mathcal{T}\mathcal{P}$.
This means that $\mathcal{P}\mathcal{T}$ is also a conjugation.

\begin{definition}\label{ddd1}
A closed densely defined linear operator $H$ in $\mathfrak{H}$ is called $\mathcal{P}\mathcal{T}$-symmetric if the relation
\begin{equation}\label{bonn3}
\mathcal{P}\mathcal{T}H=H\mathcal{P}\mathcal{T}
\end{equation}
holds on the domain $\mathcal{D}(H)$ of $H$.
\end{definition}
\begin{remark}\label{neww89}
In what follows we will often use
operator identities like
\begin{equation}\label{neww11}
XA=BX,
\end{equation}
where $A$ and $B$ are (possible) unbounded operators in a Hilbert space $\mathfrak{H}$ and  $X$ is a bounded operator in $\mathfrak{H}$.
In that case, we \emph{always assume that (\ref{neww11}) holds on $\mathcal{D}(A)$}. This means that
$X : \mathcal{D}(A)\to\mathcal{D}(B)$ and the identity $XAu=BXu$ holds for all $u\in\mathcal{D}(A)$.
If $A$ is bounded, then (\ref{neww11}) should hold on the whole $\mathfrak{H}$.
\end{remark}
In particular, relation (\ref{bonn3}) means that the conjugation $\mathcal{P}\mathcal{T}$ maps $\mathcal{D}(H)$ onto $\mathcal{D}(H)$
and $\mathcal{P}\mathcal{T}Hf=H\mathcal{P}\mathcal{T}f$ for all $f\in\mathcal{D}(H)$.

%The concept of $\mathcal{P}\mathcal{T}$-symmetry is used as a `physical analog' of the mathematical concept of adjointness  in pseudo-Hermitian %quantum mechanics \cite{BE, MO}.

\begin{lemma}\label{abbb1}
If $H$ is $\mathcal{P}\mathcal{T}$-symmetric in a Hilbert space $\mathfrak{H}$, then its adjoint operator $H^*$ is also $\mathcal{P}\mathcal{T}$-symmetric.
\end{lemma}
\emph{Proof.} Let $H$ be $\mathcal{P}\mathcal{T}$-symmetric.
It follows from (\ref{bonn1}), (\ref{bonn2}), and (\ref{bonn3}) that for all $f\in\mathcal{D}(H)$  and all $g\in\mathcal{D}(H^*)$
$$
(\mathcal{PT}Hf,g)=(H\mathcal{PT}f,g)=(\mathcal{PT}f,H^*g)=(\mathcal{PT}H^*g, f).
$$
On the other hand,
$$
(\mathcal{PT}Hf,g)=(\mathcal{T}Hf,\mathcal{P}g)=(\mathcal{TP}g, Hf)=(\mathcal{PT}g, Hf).
$$
Comparing the obtained relations, we conclude that $\mathcal{PT}g\in\mathcal{D}(H^*)$ and $\mathcal{PT}H^*g=H^*\mathcal{PT}g$ for all $g\in\mathcal{D}(H^*)$.  Hence, the adjoint operator $H^*$ is also $\mathcal{PT}$-symmetric.
Lemma \ref{abbb1} is proved. \rule{2mm}{2mm}

\begin{remark}
Recently, the similar statement was established in \cite{AT} for the particular case of space parity operator $\mathcal{P}$ and
complex conjugation $\mathcal{T}$ acting in $L_2(\mathbb{R})$.
\end{remark}

\smallskip

In general, $\mathcal{P}\mathcal{T}$-symmetric Hamiltonians admit the interpretation as
self-adjoint operators in a Krein space  $(\mathfrak{H}, [\cdot,\cdot]_{\mathcal{J}})$. However, the indefinite metric
$[\cdot,\cdot]_{\mathcal{J}}$ can not necessarily be defined by $\mathcal{J}=\mathcal{P}$  \cite{AH, MO}. In particular, for certain models \cite{GK}, the corresponding  $\mathcal{PT}$-symmetric  Hamiltonians can be interpreted as self-adjoint operators in Krein spaces
$(\mathfrak{H}, [\cdot,\cdot]_{\mathcal{J}})$  with involutions $\mathcal{J}$ constructed in terms of complex Clifford
algebras.

Let $\mathcal{P}$ and $\mathcal{R}$ be anti-commuting unitary involutions in $\mathfrak{H}$, that is,
\begin{equation}\label{es545}
\mathcal{P}\mathcal{R}=-\mathcal{R}\mathcal{P}.
\end{equation}
The operators $\mathcal{P}$ and $\mathcal{R}$ can be interpreted as generating elements of the complex Clifford algebra ${\mathcal C}l_2(\mathcal{P},\mathcal{R}):=\mbox{span}\{I, \mathcal{P}, \mathcal{R}, i\mathcal{PR}\}$ \cite{GK,LO}.

The operators $I, \mathcal{P}, \mathcal{R}$, and $i\mathcal{PR}$ are linearly independent (as a consequence of (\ref{es545})) and every operator $\mathcal{J}\in{\mathcal C}l_2(\mathcal{P},\mathcal{R})$ can thus be presented as
\begin{equation}\label{pp1}
\mathcal{J}=\alpha_0{I}+\alpha_{1}\mathcal{P}+\alpha_{2}\mathcal{R}+\alpha_{3}i\mathcal{RP}, \qquad \alpha_j\in\mathbb{C}.
\end{equation}

An operator $\mathcal{J}$ defined by (\ref{pp1}) is a non-trivial unitary involution in
$\mathfrak{H}$ (that is, $\mathcal{J}\not={\pm}I$) if and only if
\begin{equation}\label{es100}
\mathcal{J}=\alpha_{1}\mathcal{P}+\alpha_{2}\mathcal{R}+\alpha_{3}i\mathcal{RP},
\end{equation}
where $\alpha_j\in\mathbb{R}$ and $\alpha_1^2+\alpha_2^2+\alpha_3^2=1$.
The formula (\ref{es100}) establishes a one-to-one correspondence between the set of non-trivial unitary involutions $\mathcal{J}$ in
${\mathcal C}l_2(\mathcal{P},\mathcal{R})$ and vectors $\vec{\alpha}=(\alpha_1,\alpha_2,\alpha_3)$ of the unit sphere $\mathbb{S}^2$ in $\mathbb{R}^3$.

In what follows we suppose that the generators $\mathcal{P}$ and $\mathcal{R}$ of the
Clifford algebra ${\mathcal C}l_2(\mathcal{P},\mathcal{R})$ \emph{commute with the conjugation} $\mathcal{T}$:
\begin{equation}\label{bonn9}
\mathcal{P}\mathcal{T}=\mathcal{T}\mathcal{P}, \qquad \mathcal{R}\mathcal{T}=\mathcal{T}\mathcal{R}.
\end{equation}
\begin{lemma}\label{bonn8}
A non-trivial unitary involution $\mathcal{J}\in{\mathcal C}l_2(\mathcal{P},\mathcal{R})$ is $\mathcal{PT}$-symmetric if and only if
there exists $\xi\in[0,2\pi)$ such that
\begin{equation}\label{bonn14}
\mathcal{J}\equiv\mathcal{P}_\xi=e^{i\xi\mathcal{R}}\mathcal{P},
\end{equation}
where $e^{i\xi\mathcal{R}}$ is defined by the norm convergent series $\displaystyle{\sum_{n=0}^{\infty}\frac{i^n}{n!}\xi^n\mathcal{R}^n}$ in $\mathfrak{H}$. % $\mathcal{R}$ being bounded in $\mathfrak{H}$.
\end{lemma}

\emph{Proof.} If $\mathcal{P}_\xi=e^{i\xi\mathcal{R}}\mathcal{P}$, then $\mathcal{P}_\xi^*=\mathcal{P}e^{-i\xi\mathcal{R}}=e^{-i\xi\mathcal{R}}\mathcal{P}=\mathcal{P}_\xi$, where
the second equality holds due to (\ref{es545}) and
$$
\mathcal{P}_\xi^2=e^{i\xi\mathcal{R}}\mathcal{P}e^{i\xi\mathcal{R}}\mathcal{P}=e^{i\xi\mathcal{R}}e^{-i\xi\mathcal{R}}\mathcal{P}^2=I
$$
(the second equality in the latter equation holds for the same reason as before).
Hence, $\mathcal{P}_\xi$ is a unitary involution.

Conversely, a non-trivial unitary involution $\mathcal{J}\in{\mathcal C}l_2(\mathcal{P},\mathcal{R})$ is defined by (\ref{es100}).
Applying the operator $\mathcal{P}\mathcal{T}$ to both sides of (\ref{es100}) and using (\ref{es545}) and (\ref{bonn9}),
we obtain that the operator $\mathcal{J}$ is $\mathcal{P}\mathcal{T}$-symmetric if and only if $\alpha_2=0$, that is
$\mathcal{J}=\alpha_{1}\mathcal{P}+\alpha_{3}i\mathcal{RP}$, where $\alpha_1^2+\alpha_3^2=1$.

Denote $\alpha_1=\cos\xi$ and $\alpha_3=\sin\xi$, where $\xi\in[0,2\pi)$. Then
$$
\mathcal{J}=[\cos\xi]\mathcal{P}+i[\sin\xi]\mathcal{RP}=([\cos\xi]{I}+i[\sin\xi]\mathcal{R})\mathcal{P}=e^{i\xi\mathcal{R}}\mathcal{P}.
$$
Lemma \ref{bonn8} is proved. \rule{2mm}{2mm}

In many cases, a $\mathcal{PT}$-symmetric operator $H$ can be realized as a self-adjoint operator
in a Krein space $(\mathfrak{H}, [\cdot,\cdot]_{\mathcal{J}})$ for a special choice of unitary involution $\mathcal{J}\in{\mathcal C}l_2(\mathcal{P},\mathcal{R})$. The next result shows that we do not need to check this property for all $\mathcal{J}$ from
the general formula (\ref{es100}).
\begin{lemma}\label{esse1}
If a $\mathcal{PT}$-symmetric operator $H$ is self-adjoint
in a Krein space $(\mathfrak{H}, [\cdot,\cdot]_{\mathcal{J}})$ with unitary involution $\mathcal{J}$
defined by (\ref{es100}), then $H$ will also be self-adjoint in the Krein space $(\mathfrak{H}, [\cdot,\cdot]_{\mathcal{P}_\xi})$ for a certain choice of ${\mathcal{P}_\xi}$  defined by (\ref{bonn14}).
\end{lemma}
\emph{Proof.} The self-adjointness of $H$ in the Krein space $(\mathfrak{H}, [\cdot,\cdot]_{\mathcal{J}})$ is equivalent to the following  identity \cite{AZ}:
\begin{equation}\label{bonn21b}
{\mathcal{J}}Hf=H^*{\mathcal{J}}f, \qquad \forall{f}\in\mathcal{D}(H),
\end{equation}
where $H^*$ is the adjoint of $H$ in the Hilbert space $\mathfrak{H}$.

According to Lemma \ref{abbb1}, $H^*$ is $\mathcal{PT}$-symmetric, that is $\mathcal{PT}H^*=H^*\mathcal{PT}$ holds on $\mathcal{D}(H^*)$. Then, applying the bounded operator $\mathcal{PT}$ to both sides of (\ref{bonn21b}) and taking (\ref{es100}), (\ref{bonn9}) into account, we obtain that
\begin{equation}\label{assa1}
\widehat{{\mathcal{J}}}H\mathcal{PT}f=H^*\widehat{{\mathcal{J}}}\mathcal{PT}f, \qquad \forall{f}\in\mathcal{D}(H),
\end{equation}
where $\widehat{\mathcal{J}}=\alpha_{1}\mathcal{P}-\alpha_{2}\mathcal{R}+\alpha_{3}i\mathcal{RP}$.

Since $H$ is $\mathcal{PT}$-symmetric, the conjugation $\mathcal{PT}$ maps $\mathcal{D}(H)$ \emph{onto} $\mathcal{D}(H)$.
Therefore, (\ref{assa1}) can be rewritten as
$$
\widehat{{\mathcal{J}}}Hf=H^*\widehat{{\mathcal{J}}}f, \qquad \forall{f}\in\mathcal{D}(H).
$$
Summing the obtained relation with (\ref{bonn21b}) and recalling that $\mathcal{J}$
is defined by (\ref{es100}), we obtain $(\alpha_1\mathcal{P}+\alpha_3{i}\mathcal{RP})Hf=H^*(\alpha_1\mathcal{P}+\alpha_3{i}\mathcal{RP})f$ or
$$
{\mathcal{P}_\xi}Hf=H^*{\mathcal{P}_\xi}f, \qquad \forall{f}\in\mathcal{D}(H),
$$
where
$$
{\mathcal{P}_\xi}=\frac{\alpha_1}{\sqrt{1-\alpha_2^2}}\mathcal{P}+\frac{\alpha_3}{\sqrt{1-\alpha_2^2}}{i}\mathcal{RP}=([\cos\xi]{I}+i[\sin\xi]\mathcal{R})\mathcal{P}=e^{i\xi\mathcal{R}}\mathcal{P}.
$$
Therefore, $H$ is self-adjoint in the Krein space $(\mathfrak{H}, [\cdot,\cdot]_{\mathcal{P}_\xi})$.
Lemma \ref{esse1} is proved. \rule{2mm}{2mm}

\subsection{The operators $\mathcal{C}$.}
Showing that a $\mathcal{PT}$-symmetric operator $H$ turns out to be self-adjoint in some Krein space
$(\mathfrak{H}, [\cdot,\cdot]_{\mathcal{P}_\xi})$ is mathematically significant, but we have still to show
that $H$ can serve as an Hamiltonian for quantum mechanics.  To do so one must demonstrate that  $H$ is self-adjoint on a \emph{Hilbert space}
(not on a Krein space!).
This problem can be overcome for a $\mathcal{PT}$-symmetric operator $H$ by finding a new symmetry represented by a linear operator $\mathcal{C}$, which commutes with both the Hamiltonian $H$ and the $\mathcal{PT}$ operator. More precisely, suppose we can find \emph{a bounded operator} $\mathcal{C}$ ($\mathcal{C}\not={\pm}I$) in $\mathfrak{H}$  obeying the following three algebraic equations:
\begin{equation}\label{usa6}
\mathcal{C}^2=I, \qquad \mathcal{C}\mathcal{PT}=\mathcal{PT}\mathcal{C}, \qquad \mathcal{C}H=H\mathcal{C}.
\end{equation}
The third relation in (\ref{usa6}) should hold on $\mathcal{D}(H)$, see Remark \ref{neww89}.

We will say that $H$ \emph{possesses the property of $\mathcal{C}$-symmetry} if relations (\ref{usa6}) are true.

We are aiming now to analyze conditions (\ref{usa6}) assuming additionally that $\mathcal{C}\in{\mathcal C}l_2(\mathcal{P},\mathcal{R})$.
\begin{lemma}\label{ll2}
An operator $\mathcal{C}\in{\mathcal C}l_2(\mathcal{P},\mathcal{R})$ \ ($\mathcal{C}\not={\pm}I$) satisfies the conditions $\mathcal{C}^2=I$ and   $\mathcal{C}\mathcal{PT}=\mathcal{PT}\mathcal{C}$ if and only if there exist $\chi\in\mathbb{R}$ and $\xi\in[0,2\pi)$ such that
 \begin{equation}\label{es6}
 \mathcal{C}=e^{\chi{i}\mathcal{R}\mathcal{P}_\xi}\mathcal{P}_\xi
\end{equation}
(where the exponential is again defined by the corresponding norm convergent power series).
\end{lemma}
\emph{Proof.} The operator $\mathcal{P}_\xi$ is $\mathcal{PT}$-symmetric by Lemma \ref{bonn8}
and $\mathcal{PT}{i}\mathcal{R}={i}\mathcal{R}\mathcal{PT}$ due to (\ref{es545}) and (\ref{bonn9}).
Therefore, $\mathcal{PT}\chi{i}\mathcal{R}\mathcal{P}_\xi=\chi{i}\mathcal{R}\mathcal{P}_\xi\mathcal{PT}$ and
$\mathcal{C}=e^{\chi{i}\mathcal{R}\mathcal{P}_\xi}\mathcal{P}_\xi$ is $\mathcal{PT}$-symmetric.

It follows from (\ref{es545}) and (\ref{bonn14}) that $\mathcal{P}_\xi{\mathcal R}=-{\mathcal R}\mathcal{P}_\xi$. Hence,
$$
\mathcal{C}^2=e^{\chi{i}\mathcal{R}\mathcal{P}_\xi}\mathcal{P}_{\xi}e^{\chi{i}\mathcal{R}\mathcal{P}_\xi}\mathcal{P}_\xi=e^{\chi{i}\mathcal{R}\mathcal{P}_\xi}e^{-\chi{i}\mathcal{R}\mathcal{P}_\xi}\mathcal{P}_\xi^2=I.
$$

Conversely, according to (\ref{pp1}), an operator $\mathcal{C}\in{\mathcal C}l_2(\mathcal{P},\mathcal{R})$ has the form
\begin{equation}\label{bonn20}
\mathcal{C}=\alpha_0{I}+\alpha_{1}\mathcal{P}+\alpha_{2}\mathcal{R}+\alpha_{3}i\mathcal{RP}, \qquad \alpha_j\in\mathbb{C}.
\end{equation}

Using (\ref{es545}) it is easy to verify that the relation $\mathcal{C}^2=I$ \ ($\mathcal{C}\not={\pm}I$) is equivalent to the
conditions $\alpha_0=0$ and $\alpha_1^2+\alpha_2^2+\alpha_3^2=1$. Furthermore, the additional condition of $\mathcal{PT}$-symmetry
$\mathcal{C}\mathcal{PT}=\mathcal{PT}\mathcal{C}$ and (\ref{bonn9}) mean that
$\alpha_1=\overline{\alpha}_1, \ \alpha_2=-\overline{\alpha}_2$, and  $\alpha_3=\overline{\alpha}_3$.
Set  $\alpha_2'={i}\alpha_2$. Then the coefficients  $\alpha_1, \alpha_2', \alpha_3$ are real and  $\alpha_1^2+\alpha_3^2-\alpha_2'^2=1$.

Denote $\alpha_1^2+\alpha_3^2=\cosh^2\chi$ and $\alpha_2'^2=\sinh^2\chi$, where $\chi\in\mathbb{R}$. Then
$$
\alpha_1=\cos\xi\cosh\chi, \quad \alpha_3=\sin\xi\cosh\chi, \quad \alpha_2=i\sinh\chi
$$
for some $\xi\in[0,2\pi)$. Substituting the obtained expressions into (\ref{bonn20}) and using (\ref{bonn14}), we obtain
$$
\mathcal{C}=[\cosh\chi]\mathcal{P}_\xi+i[\sinh\chi]\mathcal{R}=([\cosh\chi]I+[\sinh\chi]i\mathcal{R}\mathcal{P}_\xi)\mathcal{P}_\xi=e^{\chi{i}\mathcal{R}\mathcal{P}_\xi}\mathcal{P}_\xi.
$$
Lemma \ref{ll2} is proved. \rule{2mm}{2mm}

Let us assume that a $\mathcal{PT}$-symmetric operator $H$ can be interpreted as a self-adjoint operator in the Krein space
$(\mathfrak{H}, [\cdot,\cdot]_{\mathcal{P}_\xi})$. This is equivalent to
the identity:
\begin{equation}\label{bonn21}
{\mathcal{P}_\xi}Hf=H^*{\mathcal{P}_\xi}f, \qquad \forall{f}\in\mathcal{D}(H),
\end{equation}
where the adjoint $H^*$ is understood in the sense of the initial inner product $(\cdot,\cdot)$ in $\mathfrak{H}$ \cite{AZ}.
In this case the commutation relation
$$
H\mathcal{C}f={\mathcal C}Hf, \qquad \forall{f}\in\mathcal{D}(H),
$$
where  $\mathcal{C}=e^{\chi{i}\mathcal{R}\mathcal{P}_\xi}\mathcal{P}_\xi$  (see (\ref{es6})),
means that $H$ can be realized as  \emph{a self-adjoint operator in
the Hilbert space} $\mathfrak{H}$ endowed with the new inner product \cite{AKG}
\begin{equation}\label{bonn22}
(\cdot,\cdot)_{\mathcal{C}}=[{\mathcal{C}}\cdot,\cdot]_{\mathcal{P}_\xi}=({\mathcal{P}_\xi}{\mathcal{C}}\cdot,\cdot)=(e^{-\chi{i}\mathcal{R}\mathcal{P}_\xi}\cdot,\cdot).
\end{equation}

Since ${i}\mathcal{R}\mathcal{P}_\xi$ is a bounded self-adjoint operator in $\mathfrak{H}$, the operator $e^{-\chi{i}\mathcal{R}\mathcal{P}_\xi}$ has the bounded inverse $e^{\chi{i}\mathcal{R}\mathcal{P}_\xi}$. Therefore,
the inner product $(\cdot,\cdot)_{\mathcal{C}}$ is equivalent to the initial inner product $(\cdot,\cdot)$ in $L_2(\mathbb{R})$ .

\subsection{Description of $\mathcal{PT}$-symmetric extensions.}
Let $S$ be a closed densely defined nonnegative symmetric operator  in the Hilbert space $\mathfrak{H}$.
In what follows \emph{we suppose that $S$ is $\mathcal{PT}$-symmetric}, that is, the identity $\mathcal{PT}Sf=S\mathcal{PT}f$
holds for all $f\in\mathcal{D}(S)$.  According to Lemma \ref{abbb1}, the
adjoint operator $S^*$ is also $\mathcal{PT}$-symmetric, that is, $\mathcal{PT}S^*g=S^*\mathcal{PT}g$ holds for all
$g\in\mathcal{D}(S^*)$.

\begin{lemma}\label{bonn30}
If a nonnegative symmetric operator $S$ is $\mathcal{PT}$-symmetric, then its Friedrichs extension $H_\mu$ is also
$\mathcal{PT}$-symmetric.
\end{lemma}
\emph{Proof.} Since the operators $S$ and $S^*$  are $\mathcal{PT}$-symmetric and  $S\subset{H_\mu}\subset{S^*}$, the property of
$\mathcal{PT}$-symmetry of $H_\mu$:
$$
\mathcal{PT}H_\mu{f}=H_\mu\mathcal{PT}f, \qquad \forall{f}\in\mathcal{D}(H_\mu)
$$
is equivalent to the relation
\begin{equation}\label{bonn45}
\mathcal{PT} : \mathcal{D}(H_\mu) \to \mathcal{D}(H_\mu).
\end{equation}

It is well known \cite{DS} that $\mathcal{D}(H_\mu)=\mathcal{D}_0\cap\mathcal{D}(S^*)$, where $\mathcal{D}_0\subset\mathfrak{H}$ is the closure of
$\mathcal{D}(S)$ with respect to the norm $\|\cdot\|^2_0=((S+I)\cdot,\cdot)$.

Since $\mathcal{PT} : \mathcal{D}(S) \to \mathcal{D}(S)$ (due to $\mathcal{PT}$-symmetry of $S$) and
$$
\|\mathcal{PT}f\|^2_0=((S+I)\mathcal{PT}f,\mathcal{PT}f)=(f,(S+I)f)=\|f\|^2_0
$$
the set $\mathcal{D}_0$ is invariant with respect to $\mathcal{PT}$. On the other hand,
$\mathcal{PT} : \mathcal{D}(S^*) \to \mathcal{D}(S^*)$ (due to $\mathcal{PT}$-symmetry of $S^*$).
Combining the obtained relations we deduce that relation (\ref{bonn45}) holds.
Lemma \ref{bonn30} is proved. \rule{2mm}{2mm}

Denote ${\mathcal H}=\ker(S^{*}+I)$. Then $\mathcal{D}(S^{*})={\mathcal D}(H_{\mu})\dot{+}{\mathcal H}$ and, hence,
an arbitrary function $f\in\mathcal{D}(S^{*})$ is uniquely decomposed as follows:
\begin{equation}\label{bonn41}
f=u+h, \qquad u\in{\mathcal{D}(H_{\mu})}, \quad h\in{\mathcal H}.
\end{equation}

The decomposition (\ref{bonn41}) allows one to define the linear mappings $\Gamma_{0}$ and $\Gamma_{1}$ from
$\mathcal{D}(S^{*})$ into ${\mathcal H}$:
  \begin{equation}\label{eee7}
   \Gamma_{0}f=h, \quad  \Gamma_{1}f=P_{\mathcal H}(H_{\mu}+{I})u, \qquad f\in\mathcal{D}(S^{*}), \quad u\in{\mathcal{D}(H_{\mu})},
    \end{equation}
where $P_{\mathcal H}$ is the orthogonal projector onto the subspace ${\mathcal H}$ in $\mathfrak H$.

 The triple $({\mathcal H},\Gamma_{0},\Gamma_{1})$ is called the positive boundary triplet of $S^*$
    associated with the Friedrichs extension $H_{\mu}$ \cite{Gor}.

 It follows from (\ref{bonn41}) and (\ref{eee7}) that an arbitrary intermediate extension $H$ of $S$ (that is, $S\subset{H}\subset{S^*}$)
 with $-1\in\rho(H)$ (where $\rho(H)$ denotes the resolvent set of $H$) is defined as follows:
 \begin{equation}\label{eee8}
  H=S^*\upharpoonright_{\mathcal{D}(H)}, \qquad  \mathcal{D}(H)=\{f\in{\mathcal{D}(S^{*})} \ :  \  T\Gamma_{1}f=\Gamma_{0}f\},
    \end{equation}
  where $T$ is a bounded operator in ${\mathcal H}$.

  The formulas (\ref{eee7}) enable one to establish the following
  relationship between $H$ and $T$ in (\ref{eee8}):
  \begin{equation}\label{bonn42}
  T=[(H+{I})^{-1}-(H_{\mu}+{I})^{-1}]\upharpoonright_{\mathcal H}.
  \end{equation}

 \begin{lemma}\label{bonn43}
  Let $H$ be defined by (\ref{eee8}). Then $H$ is $\mathcal{PT}$-symmetric if and only if
  the bounded operator $T$ is $\mathcal{PT}$-symmetric.
 \end{lemma}
 \emph{Proof}. Since $S^*$ commutes with $\mathcal{PT}$ (in the sense specified before Lemma \ref{bonn30}), the subspace $\mathcal{H}$ is invariant with respect to $\mathcal{PT}$,
 that is, $\mathcal{PT}\mathcal{H}=\mathcal{H}$. This means that the restriction of $\mathcal{PT}$ onto $\mathcal{H}$ gives rise to
 a conjugation operator in $\mathcal{H}$.  Taking into account Lemma \ref{bonn30} and decomposition (\ref{bonn41}), we conclude that
 the mappings $\Gamma_0$, $\Gamma_1$ defined by (\ref{eee7}) commute with $\mathcal{PT}$:
 \begin{equation}\label{bonn44}
\mathcal{PT}\Gamma_0=\Gamma_0\mathcal{PT}, \qquad \mathcal{PT}\Gamma_1=\Gamma_1\mathcal{PT}.
\end{equation}

Let $T$ be $\mathcal{PT}$-symmetric, that is $T\mathcal{PT}=\mathcal{PT}T$. The operator $H$ defined by (\ref{eee8}) is
$\mathcal{PT}$-symmetric if and only if
$\mathcal{PT} : \mathcal{D}(H) \to \mathcal{D}(H)$, c.f. (\ref{bonn45}).  This relation is equivalent
to the identity $T\mathcal{PT}=\mathcal{PT}T$, in view of (\ref{bonn44}). Thus $H$ is $\mathcal{PT}$-symmetric.

Conversely, if $H$ is $\mathcal{PT}$-symmetric, then $T$ is $\mathcal{PT}$-symmetric due to Lemma \ref{bonn30} and relation (\ref{bonn42}).
Lemma \ref{bonn43} is proved. \rule{2mm}{2mm}

\smallskip

In what follows we suppose that $S$ \emph{commutes with all elements} of the Clifford algebra ${\mathcal C}l_2(\mathcal{P}, \mathcal{R})$ or, that is equivalent, the operator identities
\begin{equation}\label{es2}
S\mathcal{P}={\mathcal{P}}S, \qquad S\mathcal{R}=\mathcal{R}S
\end{equation}
hold on $\mathcal{D}(S)$. Considering the adjoint operators of both parts of (\ref{es2})  and remembering that $\mathcal{P}$ and $\mathcal{R}$ are unitary involutions, we obtain that the following identities hold on $\mathcal{D}(S^*)$:
\begin{equation}\label{sas1b}
(S\mathcal{P})^*={\mathcal{P}}S^*=({\mathcal{P}}S)^*=S^*\mathcal{P}, \quad (S\mathcal{R})^*={\mathcal{R}}S^*=({\mathcal{R}}S)^*=S^*\mathcal{R}.
\end{equation}
Hence, the restrictions of $\mathcal{P}$ and $\mathcal{R}$ onto $\mathcal{H}$ determine unitary involutions in $\mathcal{H}$ (we will keep the same notations $\mathcal{P}$ and  $\mathcal{R}$ for these restrictions).

It follows from (\ref{es2}) that the Friedrichs extension $H_\mu$ of $S$ commutes with $\mathcal{P}$ and $\mathcal{R}$.
This allows one to establish an analog of (\ref{bonn44}):
\begin{equation}\label{bonn44b}
{\mathcal P}\Gamma_j=\Gamma_j{\mathcal P}, \qquad
{\mathcal R}\Gamma_j=\Gamma_j{\mathcal R}, \qquad j=0,1,
\end{equation}
where the restrictions of ${\mathcal P}$ and ${\mathcal R}$
onto $\mathcal{H}$ are anti-commuting unitary involutions in the Hilbert space $\mathcal{H}$.
Therefore, the formulas (\ref{pp1}) and (\ref{bonn44b})
\emph{ensure a bijective correspondence between
elements of the initial Clifford algebra ${\mathcal C}l_2({\mathcal P}, \mathcal{R})$ and then
restrictions onto the auxiliary space $\mathcal{H}$}.
In particular, for every $\mathcal{P}_\xi$ defined by (\ref{bonn14}),
we have
\begin{equation}\label{es45}
\mathcal{P}_\xi\Gamma_j=\Gamma_j\mathcal{P}_\xi,  \qquad  j=0,1.
\end{equation}

\begin{lemma}\label{es16}
Let $H$ be defined by (\ref{eee8}). Then $H$
is a self-adjoint operator in the Krein space $(\mathfrak{H}, [\cdot,\cdot]_{\mathcal{P}_\xi})$ if and only if
the bounded operator $T$ in (\ref{eee8}) is self-adjoint in the Krein space
$(\mathcal{H}, [\cdot,\cdot]_{\mathcal{P}_\xi})$.
\end{lemma}
\emph{Proof.} By virtue of (\ref{sas1b}), the operator $S^*$ commutes with the generators $\mathcal{P}$ and $\mathcal{R}$ of the
Clifford algebra $\mathcal{C}l_2(\mathcal{P}, \mathcal{R})$. This means that $S^*$ commutes with
an arbitrary operator from $\mathcal{C}l_2(\mathcal{P}, \mathcal{R})$. In particular, the relation
\begin{equation}\label{sas2b}
{\mathcal{P}_\xi}S^*g=S^*{\mathcal{P}_\xi}g, \qquad \forall{g}\in\mathcal{D}(S^*)
\end{equation}
holds, where $\mathcal{P}_\xi$ is defined by (\ref{bonn14}).

The self-adjointness of $H$ in the Krein space $(\mathfrak{H}, [\cdot,\cdot]_{\mathcal{P}_\xi})$ is equivalent to the
identity (\ref{bonn21}), which is equivalent to the relation
\begin{equation}\label{bonn50}
{\mathcal{P}_\xi} : \mathcal{D}(H) \to \mathcal{D}(H^*),
\end{equation}
due to (\ref{sas2b}) and the relation $S\subset{H}\subset{S^*}$.

Since $H$ is determined by (\ref{eee8}), the corresponding formula for $H^*$ takes the form
 \begin{equation}\label{eee8bb}
  H^*=S^*\upharpoonright_{\mathcal{D}(H^*)}, \quad  \mathcal{D}(H^*)=\{f\in{\mathcal{D}(S^{*})} \ :  \  T^*\Gamma_{1}f=\Gamma_{0}f\},
    \end{equation}
where $T^*$ is the adjoint of $T$ in the Hilbert space $\mathcal{H}$ \cite{Gor}.
Combining (\ref{eee8}) with (\ref{eee8bb}) and taking into account (\ref{es45}), we deduce that the relation (\ref{bonn50}) is equivalent
to the condition $T{\mathcal{P}_\xi}={\mathcal{P}_\xi}T^*$, which means that $T$ is self-adjoint in the Krein space
$(\mathcal{H}, [\cdot,\cdot]_{\mathcal{P}_\xi})$. Lemma \ref{es16} is proved.
\rule{2mm}{2mm}
\subsection{The case of deficiency indices $<2,2>$.}
\begin{theorem}\label{es16b}
If a nonnegative symmetric operator $S$ has deficiency indices $<2,2>$ in
$\mathfrak{H}$, then its arbitrary $\mathcal{PT}$-symmetric extension $H$ defined by (\ref{eee8})
can be interpreted as a self-adjoint operator in the Krein space $(\mathfrak{H}, [\cdot,\cdot]_{\mathcal{P}_\xi})$ for a certain choice
of $\xi\in[0,2\pi)$.
\end{theorem}
\emph{Proof.}
According to Lemma \ref{bonn43}, $\mathcal{PT}$-symmetric extensions $H$ of $S$ defined by (\ref{eee8})
are described by $\mathcal{PT}$-symmetric operators $T$ acting in $\mathcal{H}=\ker(S^*+I)$.

Since $S$ has deficiency indices $<2,2>$, the dimension of $\mathcal{H}$ is equal $2$.
This means that the Clifford algebra ${\mathcal C}l_2(\mathcal{P},\mathcal{R})$  with generators $\mathcal{P}$ and $\mathcal{R}$  considered
on $\mathcal{H}$ \emph{coincides with the set of all operators acting in} $\mathcal{H}$. Therefore,
\begin{equation}\label{pp1b}
T=\alpha_0{I}+\alpha_{1}\mathcal{P}+\alpha_{2}\mathcal{R}+\alpha_{3}i\mathcal{RP}, \qquad \alpha_j\in\mathbb{C}.
\end{equation}
It follows from (\ref{bonn2b}), (\ref{bonn9}), and (\ref{pp1b}) that
$$
\mathcal{PT}T=(\overline{\alpha}_0{I}+\overline{\alpha}_{1}P-\overline{\alpha}_2\mathcal{R}+\overline{\alpha}_3i\mathcal{RP})\mathcal{PT}.
$$
Hence, $T$ is $\mathcal{PT}$-symmetric if and only if
\begin{equation}\label{bonn80}
\alpha_0=\overline{\alpha}_0, \quad \alpha_1=\overline{\alpha}_1, \quad \alpha_2=-\overline{\alpha}_2, \quad \alpha_3=\overline{\alpha}_3.
\end{equation}

Let the involution $\mathcal{P}_\xi$ be defined by (\ref{bonn14}). According to Lemma \ref{es16}, the operator $H$ is
self-adjoint in the Krein space $(\mathfrak{H}, [\cdot,\cdot]_{\mathcal{P}_\xi})$ if the operator $T$ is self-adjoint in
$(\mathcal{H}, [\cdot,\cdot]_{\mathcal{P}_\xi})$. The latter condition is equivalent to the operator identity
$T{\mathcal{P}_\xi}={\mathcal{P}_\xi}T^*$. A simple calculation shows:
$$
\begin{array}{l}
{T{\mathcal{P}_\xi}=Te^{i\xi\mathcal{R}}\mathcal{P}=T([\cos\xi]\mathcal{P}+i[\sin\xi]\mathcal{R}\mathcal{P})=[\alpha_1\cos\xi+\alpha_3\sin\xi]I+} \vspace{3mm} \\ 
{[\alpha_0\cos\xi+i\alpha_2\sin\xi]{\mathcal P}+[i\alpha_3\cos\xi-i\alpha_1\sin\xi]{\mathcal R}+[-i\alpha_2\cos\xi+i\alpha_0\sin\xi]i{\mathcal R}{\mathcal P}}
\end{array}
$$
and
$$
\begin{array}{l}
{{{\mathcal P}_\xi}T^*=e^{i\xi\mathcal{R}}\mathcal{P}T^*=([\cos\xi]\mathcal{P}+i[\sin\xi]\mathcal{R}\mathcal{P})T^*=[\overline{\alpha}_1\cos\xi+\overline{\alpha}_3\sin\xi]I+} \vspace{3mm} \\ 
{[\overline{\alpha}_0\cos\xi-i\overline{\alpha}_2\sin\xi]\mathcal{P}+[-i\overline{\alpha}_3\cos\xi+i\overline{\alpha}_1\sin\xi]{\mathcal R}+[i\overline{\alpha}_2\cos\xi+i\overline{\alpha}_0\sin\xi]i\mathcal{R}\mathcal{P}}
\end{array}
$$
Comparing the obtained expressions and taking (\ref{bonn80}) into account, we arrive at the conclusion that the $\mathcal{PT}$-symmetric operator
$T$ is self-adjoint in the Krein space  $(\mathcal{H}, [\cdot,\cdot]_{\mathcal{P}_\xi})$ if and only if
\begin{equation}\label{bonn200}
\alpha_1\sin\xi=\alpha_3\cos\xi.
\end{equation}
Thus, every $\mathcal{PT}$-symmetric operator $T$ can be interpreted as a self-adjoint operator in the Krein space $(\mathcal{H}, [\cdot,\cdot]_{\mathcal{P}_\xi})$, where the parameter $\xi$ is determined by (\ref{bonn200}).
Theorem \ref{es16b} is proved \rule{2mm}{2mm}

\begin{theorem}\label{esse8}
 Let a $\mathcal{PT}$-symmetric operator $H$ be defined by (\ref{eee8}) and let
 $H$ be self-adjoint in the Krein space $(\mathfrak{H}, [\cdot,\cdot]_{\mathcal{P}_\xi})$. Then $H$ possesses the property of $\mathcal{C}$-symmetry with $\mathcal{C}=e^{\chi{i}\mathcal{R}\mathcal{P}_\xi}\mathcal{P}_\xi$ if and only if the operator $T$ in (\ref{eee8}) has the form
\begin{equation}\label{bonn301}
T=\beta_0{I}+\beta_1\mathcal{C}, \qquad \beta_0, \ \beta_1\in\mathbb{R}.
\end{equation}
\end{theorem}
\emph{Proof.} In view of Lemma \ref{bonn43}, operators $H$ defined by (\ref{eee8}) are $\mathcal{PT}$-symmetric
if and only if the corresponding operators $T$ are $\mathcal{PT}$-symmetric.
The operators $T$ act in the space $\mathcal{H}$ and they are described completely\footnote{Since $\mathcal{H}$ has dimension $2$.} with
the help of (\ref{pp1b}).

Since $\mathcal{P}_\xi=e^{i\xi\mathcal{R}}\mathcal{P}$, the unitary involutions $\mathcal{P}_\xi$ and $\mathcal{R}$ anti-commute.
Hence, they can be considered as generating elements of the Clifford algebra ${\mathcal C}l_2(\mathcal{P},\mathcal{R})$, that is,
${\mathcal C}l_2(\mathcal{P},\mathcal{R})={\mathcal C}l_2(\mathcal{P}_\xi,\mathcal{R})$. This means that
the operator $T$ given by (\ref{pp1b}) can also be rewritten as:
\begin{equation}\label{bonn203}
T=\beta_0{I}+\beta_{1}\mathcal{P}_\xi+\beta_{2}\mathcal{R}+\beta_{3}i\mathcal{R}\mathcal{P}_\xi, \qquad \beta_j\in\mathbb{C}
\end{equation}

If $T$ corresponds to a $\mathcal{PT}$-symmetric operator $H$ in (\ref{eee8}), which is self-adjoint in the Krein space $(\mathcal{H}, [\cdot,\cdot]_{\mathcal{P}_\xi})$, then the coefficients $\alpha_j$ of $T$ satisfy (\ref{bonn80}) and (\ref{bonn200}), due to
Theorem \ref{es16b}. In that case, comparing
(\ref{pp1b}) and (\ref{bonn203}) and taking into account that
$$
\mathcal{P}_\xi=e^{i\xi\mathcal{R}}\mathcal{P}=(\cos\xi)\mathcal{P}+i(\sin\xi)\mathcal{RP},
$$
we obtain the following relationship between $\alpha_j$ and $\beta_j$:
$$
\beta_0=\alpha_0\in\mathbb{R}, \quad \beta_1=(\alpha_1\cos\xi+\alpha_3\sin\xi)\in\mathbb{R}, \quad \beta_2=\alpha_2\in{i\mathbb{R}}, \quad \beta_3=0.
$$

Thus, the $\mathcal{PT}$-symmetric operators $H$ in (\ref{eee8}), which are also self-adjoint operators in the Krein space
$(\mathcal{H}, [\cdot,\cdot]_{\mathcal{P}_\xi})$ are distinguished by the operator-parameters
 \begin{equation}\label{bonn206}
T=\beta_0{I}+\beta_{1}\mathcal{P}_\xi+\beta_{2}\mathcal{R},
\end{equation}
with $\beta_0, \beta_1\in\mathbb{R}$ and $\beta_2\in{i}\mathbb{R}$.

The operator $\mathcal{C}=e^{\chi{i}\mathcal{R}\mathcal{P}_\xi}\mathcal{P}_\xi$  has a simple presentation with respect to the unitary involutions $\mathcal{P}_\xi$ and $\mathcal{R}$:
\begin{equation}\label{bonn207}
\mathcal{C}=e^{\chi{i}\mathcal{R}\mathcal{P}_\xi}\mathcal{P}_\xi=(\cosh\chi)\mathcal{P}_\xi+i(\sinh\chi)\mathcal{R}.
\end{equation}

It follows from (\ref{bonn44b}), (\ref{es45}), and (\ref{bonn207}) that $\mathcal{C}\Gamma_j=\Gamma_j\mathcal{C}$, $j=0,1$.
Then, repeating the proof of Lemma \ref{bonn43} with the use of $\mathcal{C}$ instead of $\mathcal{PT}$,  we deduce that the relation $\mathcal{C}Hf=H\mathcal{C}f$ holds for all $f\in\mathcal{D}(H)$ if and only if the identity  $\mathcal{C}T=T\mathcal{C}$ holds on $\mathcal{H}$.

A simple calculation with the use of (\ref{bonn206}) and (\ref{bonn207}) shows that the operator identity
$\mathcal{C}T=T\mathcal{C}$ is equivalent to the condition
$\beta_2=i\beta_1\tanh\chi$. Substituting this expression into (\ref{bonn206}), we obtain
$$
T=\beta_0{I}+\frac{\beta_1}{\cosh\chi}([\cosh\chi]I+i[\sinh\chi]R\mathcal{P}_\xi)\mathcal{P}_\xi=\beta_0{I}+\frac{\beta_1}{\cosh\chi}e^{\chi{i}\mathcal{R}\mathcal{P}_\xi}\mathcal{P}_\xi,
$$
where $\beta_0, \beta_1$ are arbitrary real parameters. This leads to (\ref{bonn301}) if we set $\beta_1=\frac{\beta_1}{\cosh\chi}$.
Theorem \ref{esse8} is proved. \rule{2mm}{2mm}

\section{Elements of the Lax-Phillips scattering scheme. The self-adjoint case}
\subsection{Definition of free and perturbed evolutions.}
For developing a Lax-Phillips scattering theory for the operator-differential equation
\begin{equation}\label{bonn60}
\frac{d^2}{dt^2}u=-Hu,
\end{equation}
one should first define a free and a perturbed dynamics, which satisfy the requirements of the Lax-Phillips approach
\cite{LF}. For the operator-differential equation (\ref{bonn60}) with a nonnegative self-adjoint operator $H$ acting in a Hilbert space
$\mathfrak{H}$ this problem was analyzed in \cite{KU1,AlAn}. We briefly outline the principal points.

A symmetric operator $B$ in a Hilbert space ${\mathfrak H}$ is called \emph{simple} if it does not induce a self-adjoint
operator in any proper subspace of ${\mathfrak H}$; $B$ is called \emph{maximal symmetric} if one of its defect numbers
is equal to zero.

In what follows, without loss of generality, we assume that the defect number of $B$ vanishes in the
lower half plane $\mathbb{C}_-$, that is, $\mathcal{R}(B-z{I})=\mathfrak{H}$ for all $z\in\mathbb{C}_-=\{z\in\mathbb{C} : \textsf{Im} \ z<0\}$. In that case the simple maximal symmetric operator $B$ has the representation:
\begin{equation}\label{e14}
 B=X^{-1}i\frac{d}{dx}X, \qquad D(B)=X^{-1}\{u(x)\in{W^1_2}({\mathbb R}_+,N) \ | \ u(0)=0\},
\end{equation}
where $X$ isometrically maps $\mathfrak{H}$ onto $L_2({\mathbb R}_+,N)$
and the dimension of the Hilbert space $N$ is equal
to the nonzero defect number of $B$ \cite{AkGl}.

It follows from (\ref{e14}) that the operator
$B^2$ is unitarily equivalent to the operator $-\frac{d^2}{dx^2}$ defined on
$\{u(x)\in{W^2_2}({\mathbb R}_+,N) \ | \ u(0)=u'(0)=0\}$ in $L_2({\mathbb R}_+,N)$.
Therefore, $B^2$ is a nonnegative symmetric operator in $\mathfrak{H}$.

\begin{definition}\label{bonn70}
A self-adjoint extension $H$ of $B^2$ in the Hilbert space $\mathfrak{H}$ is called \emph{a (Lax-Phillips) unperturbed operator} if
    \begin{equation}\label{esse5}
    (Hf,f)={\Vert B^{*}f \Vert}^{2}, \qquad \forall{f}\in{\mathcal{D}(H)}.
    \end{equation}
 \end{definition}

Let $H$ be an unperturbed operator in ${\mathfrak{H}}$.
Denote by ${\mathfrak{H}}_{H}$ the completion of the domain $\mathcal{D}(H)$  with respect to the norm $\|\cdot\|^2_H:=(H\cdot, \cdot)$. In the energy space $\mathfrak{G}=\mathfrak{H}_{H}\oplus{\mathfrak{H}}$, the
equation (\ref{bonn60}) naturally defines a unitary group $W_{H}(t)$ of solutions of the Cauchy problem \cite{LF}.

Denote by  $D_-$ resp. $D_+$ the closures (in $\mathfrak{G}$) of the
sets\footnote{We present elements of $\mathfrak{G}_H$ as $\left(\begin{array}{c}u \\
v\end{array}\right)$, where $u\in\mathfrak{H}_H$ and $v\in\mathfrak{H}$.}
\begin{equation}\label{sese1}
\left\{\left(\begin{array}{c}
u \\
-iBu
\end{array}\right) \right\} \quad \mbox{resp.} \quad
\left\{\left(\begin{array}{c}
u \\
iBu
\end{array}\right) \right\}, \qquad  \forall{u}\in{D(B^2)}.
\end{equation}

The subspaces $D_-$ and $D_+$ are, respectively, \emph{incoming} and  \emph{outgoing}
subspaces for $W_{H}(t)$ in the sense that:
\begin{equation}\label{bonn62}
\begin{array}{l}
 (i) \quad  W_H(-t)D_{-}\subset{D_{-}}, \qquad  W_H(t)D_{+}\subset{D_{+}}, \qquad t\geq0; \vspace{3mm} \\
 (ii) \quad \bigcap_{t\geq 0}W_H(-t)D_{-}=\bigcap_{t\geq 0}W_H(t)D_{+}=\{0\}; \vspace{3mm} \\
 (iii) \quad D_-\oplus{D_+}=\mathfrak{G}
\end{array}
\end{equation}
(the sign $\oplus$ means the orthogonality in $\mathfrak{G}$).

The conditions (\ref{bonn62}) characterize \emph{the free evolution in the Lax-Phillips scattering framework} \cite{KU1, LF}.
Therefore, Eq. (\ref{bonn60}) with an unperturbed operator $H$ determines the Lax-Phillips free evolution $W_H(t)$.

It is easy to check that the relation (\ref{esse5}) holds for the Friedrichs extension $H_\mu=B^*B$ of $B^2$.
Therefore, $H_\mu$ is an example of unperturbed operator in the sense of Definition \ref{bonn70}.

In what follows, for definiteness, we will assume that the free evolution is determined by the
unitary group of operators $W_{H_\mu}(t)$, where $H_\mu$ is the Friedrichs extension of $B^2$.

\begin{definition}\label{ddd6}
A nonnegative self-adjoint operator ${H}$ in a Hilbert space $\widetilde{\mathfrak{H}}$ is called  \emph{perturbed} if
$\widetilde{\mathfrak{H}}$ contains ${\mathfrak{H}}$ as a subspace and
${H}$ is an extension of $B^2$.
A perturbed operator ${H}$ is called $0$-perturbed if $\widetilde{\mathfrak{H}}={\mathfrak{H}}$.
\end{definition}

Likewise, Eq. (\ref{bonn60}) with a perturbed operator $H$ on the right hand side determines a unitary group $W_{{H}}(t)$ of solutions
of the Cauchy problem in the new energy space $\mathfrak{G}_{{H}}=\mathfrak{H}_{{H}}\oplus{\widetilde{\mathfrak{H}}}$, which contains $\mathfrak{G}$ as a subspace.

The subspaces $D_\pm$ defined above belong to $\mathfrak{G}_{{H}}$ and they satisfy
the conditions (i), (ii) of (\ref{bonn62}) with respect to the group of unitary operators $W_{{H}}(t)$.
This corresponds to the case of perturbed evolution defined in the Lax-Phillips scattering scheme \cite{LF}.
Thus, Eq, (\ref{bonn60}) with perturbed operator ${H}$ determines the Lax-Phillips perturbed evolution.

\subsection{Scattering matrix. Definition and calculation.}
The existence of \emph{the same} incoming $D_-$ and outgoing $D_+$ subspaces for the unperturbed $W_{H_\mu}(t)$ and the perturbed $W_{H}(t)$ groups allows one to establish the existence of  the wave operators
$$
\Omega_{\pm}=s-\lim_{t\to\pm\infty}W_{H}(-t)W_{H_\mu}(t),
$$
which turn out to be complete if $H$ is $0$-perturbed \cite[Proposition 2.1]{AlAn}.

A unitary mapping $F$ of the energy space $\mathfrak{G}$ onto $L_2(\mathbb{R}, N)$ that defines a spectral representation of
the unperturbed group $W_{H_\mu}(t)$ can be constructed with the use of (\ref{e14}) \cite{KU1,AlAn}.

The scattering operator in the spectral representation
$$
\mathbb{S}=F\Omega_+^*\Omega_-F^{-1}
$$
can be realized as the operator of multiplication by a function $\mathbb{S}(\delta)\ (\delta\in\mathbb{R})$ whose values are contraction operators in the auxiliary space $N$.
The function $\mathbb{S}(\delta)$ is called \emph{the scattering matrix for the perturbed group} $W_{{H}}(t)$.

Since the subspaces $D_-$ and $D_+$ defined by (\ref{sese1}) are orthogonal in the energy space $\mathfrak{G}_H$,
the standard arguments \cite{LF}
lead to the conclusion that the scattering matrix $\mathbb{S}(\delta)$ is a boundary value in the sense of strong convergence of a contraction operator-valued function $\mathbb{S}(z)$ analytic in the lower half-plane.

The function $\mathbb{S}(z)$ is closely related to the perturbed operator ${H}$ and the investigation of the relationship between
$\mathbb{S}(z)$ and ${H}$ is the proper subject of Lax-Phillips scattering theory.

The common feature of the unperturbed $H_\mu$ and the perturbed ${H}$ operators is that they \emph{are extensions of
a given symmetric operator} $B^2$. This gave rise to a simple recipe for finding $\mathbb{S}(z)$ \cite{AlAn}.

In particular, if $H$ is $0$-perturbed operator in the sense of Definition \ref{ddd6},
then $H$ is a nonnegative self-adjoint extension of $B^2$ acting in $\mathfrak{H}$.
In that case, $H$ is defined by   (\ref{eee8}),
 where $T=[(H+{I})^{-1}-(H_\mu+{I})^{-1}]\upharpoonright_{\mathcal H}$
 is a bounded operator in ${\mathcal H}=\ker({B^*}^2+I)$. The dimension of ${\mathcal H}$ coincides with the dimension of
 the auxiliary space $N$ in the definition of the scattering matrix. Therefore, we can identify ${\mathcal H}$ and $N$.
\begin{theorem}[\cite{AlAn}]\label{esse3}
If ${H}$ is a $0$-perturbed operator, then the analytic continuation $\mathbb{S}(z)$ of the scattering matrix $\mathbb{S}(\delta)$ for the perturbed group $W_H(t)$ has the
form\footnote{we use the notation $\frac{1}{X}$ for the inverse operator $X^{-1}$.}
\begin{equation}\label{bonn65}
\mathbb{S}(z)=\frac{I-2(1+iz)T}{I-2(1-iz)T}, \qquad z\in\mathbb{C}_-,
\end{equation}
where $T$ is taken from (\ref{eee8}).
\end{theorem}

%The holomorphic function $\mathbb{S}(z)$ determines a perturbed self-adjoint operator ${H}$ up to equivalence. In order to formulate
%the corresponding result we recall that: (a)
%a perturbed operator ${H}$ is called \emph{minimal} if any subspace of $\widetilde{\mathfrak{H}}\ominus{\mathfrak{H}}$  that reduces ${H}$ is %trivial; \ (b)
%   minimal perturbed extensions ${H}_1$ and ${H}_2$ of $B^2$ acting in spaces $\widetilde{\mathfrak{H}}_1$ and $\widetilde{\mathfrak{H}}_2$, %respectively, are called \emph{equivalent} if there exists an isometric mapping $X$ of $\widetilde{\mathfrak{H}}_1$ onto %$\widetilde{\mathfrak{H}}_2$ such that ${H}_2=X{H}_1X^{-1}$
%and $Xf=f$ for all $f\in\mathfrak{H}$.
%
%\begin{theorem}[\cite{AlAn}]
%The analytic continuation ${\mathbb{S}}(z)$ of the scattering matrix uniquely determines a minimal
%perturbed self-adjoint extension ${H}$ of $B^2$ up to equivalence.
%\end{theorem}

\section{Elements of the Lax-Phillips scattering scheme. The $\mathcal{PT}$-symmetric case.}
\subsection{Preliminaries.}
In what follows we assume that a simple maximal symmetric operator $B$ \emph{commutes with elements
of the Clifford algebra $Cl_2(\mathcal{P}, \mathcal{R})$ and anti-commutes with $\mathcal{T}$}. This means that
the following operator identities hold on $\mathcal{D}(B)$ (see Remark \ref{neww89}):
\begin{equation}\label{zvit1}
\mathcal{P}B=B\mathcal{P}, \quad \mathcal{R}B=B\mathcal{R}, \quad \mathcal{T}B=-B\mathcal{T}.
\end{equation}
It is easy to see that \emph{the analogous commutation properties} remain true for the adjoint operator $B^*$:
\begin{equation}\label{zvit2}
\mathcal{P}B^*=B^*\mathcal{P}, \quad \mathcal{R}B^*=B^*\mathcal{R}, \quad \mathcal{T}B^*=-B^*\mathcal{T}.
\end{equation}
It follows from (\ref{zvit1}) and (\ref{zvit2}) that
\begin{equation}\label{zvit3}
\mathcal{P}X=X\mathcal{P}, \quad \mathcal{R}X=X\mathcal{R}, \quad \mathcal{T}X=X\mathcal{T}, \quad X\in\{B^2, {B^*}^2, H_\mu\}.
\end{equation}

Consider the Hilbert space $\mathfrak{H}$ endowed
with the new inner product $(\cdot,\cdot)_{\mathcal{C}}=(\mathcal{P}_\xi\mathcal{C}\cdot, \cdot)$ defined by (\ref{bonn22}).
Since $\mathcal{C}$ resp. $\mathcal{P}_\xi$ are defined as:  $\mathcal{C}=e^{\chi{i}\mathcal{R}\mathcal{P}_\xi}\mathcal{P}_\xi$ resp. $\mathcal{P}_\xi=e^{i\xi\mathcal{R}}\mathcal{P}$, these operators belong to ${Cl_2(\mathcal{P}, \mathcal{R})}$ for all
$\chi\in\mathbb{R}$ and $\xi\in[0,2\pi)$. Therefore, the operators $B$ and $B^*$ commute with the operator $\mathcal{P}_\xi\mathcal{C}=e^{-\chi{i}\mathcal{R}\mathcal{P}_\xi}$ due to (\ref{zvit1}) and (\ref{zvit2}).
Taking into account these commutation properties in the definition
(\ref{bonn22}) of new inner product $(\cdot,\cdot)_{\mathcal{C}}$ we derive that
$B$ \emph{keeps being a simple maximal symmetric operator\footnote{see the corresponding definition in Subsection 3.1} with respect to the inner product $(\cdot,\cdot)_{\mathcal{C}}$ and its adjoint with respect to $(\cdot,\cdot)_{\mathcal{C}}$ coincides with the adjoint $B^*$ with respect to $(\cdot,\cdot)$.}

This means that the Friedrichs extension $H_\mu=B^*B$ of $B^2$ does not depend on $(\cdot,\cdot)_{\mathcal{C}}$ and $H_\mu$ is a Lax-Phillips unperturbed operator in the sense of Definition \ref{bonn70} \emph{for any choice of inner product} $(\cdot,\cdot)_{\mathcal{C}}$ defined by
(\ref{bonn22}).

Let an extension ${H}$ of $B^2$ be a self-adjoint operator in the Krein space $(\mathfrak{H}, [\cdot,\cdot]_{\mathcal{P}_\xi})$.
Assume also that the spectrum of ${H}$ is nonnegative and, \emph{for a certain choice of}  $\mathcal{C}=e^{\chi{i}\mathcal{R}\mathcal{P}_\xi}\mathcal{P}_\xi$, the operator $H$ commutes with $\mathcal{C}$, that is,
the identity $H\mathcal{C}=\mathcal{C}H$ holds on $\mathcal{D}(H)$. Then the operator $H$ turns out to be self-adjoint in the Hilbert space $\mathfrak{H}$
with the inner product $(\cdot,\cdot)_{\mathcal{C}}$ (see the end of subsection 2.2).
Furthermore, this self-adjoint operator is nonnegative since the
spectrum of $H$ is nonnegative. This means that $H$ is a $0$-perturbed operator in the sense of
Definition \ref{ddd6}.

Summing up: \emph{the pair of operators $H_\mu$ and $H$ turns out to be unperturbed and $0$-perturbed operators in the Lax-Phillips scheme, respectively, if we endow the Hilbert space $\mathfrak{H}$ with new inner product $(\cdot,\cdot)_{\mathcal{C}}$, which is equivalent to
the initial one $(\cdot,\cdot)$}. This means that we can use Theorem \ref{esse3} for this pair of operators and the corresponding
scattering matrix $\mathbb{S}(\delta)$ has an analytical continuation $\mathbb{S}(z)$ in the lower half-plane, which is defined by
(\ref{bonn65}).

The values of the function $\mathbb{S}(z)$ are operators in the auxiliary space $N=\mathcal{H}$. These operators are contraction operators with respect to the inner product $(\cdot,\cdot)_{\mathcal{C}}$. This property is not preserved if we consider $\mathbb{S}(z)$ with respect
to the original inner product $(\cdot,\cdot)$. The same is true for the unitary property of the scattering matrix
\begin{equation}\label{bonn65b}
\mathbb{S}(\delta)=\frac{I-2(1+i\delta)T}{I-2(1-i\delta)T}, \quad \delta\in\mathbb{R},
\end{equation}
which takes place with respect to inner product $(\cdot,\cdot)_{\mathcal{C}}$ only.

Since $\mathcal{C}$ depends on the choice of $H$, different inner products $(\cdot,\cdot)_{\mathcal{C}}$ in $\mathfrak{H}$
have to be used for various operators $H$. From this point of view, it looks reasonable to consider $\mathbb{S}(\delta)$ with respect to the
initial inner product $(\cdot,\cdot)$ and to describe the changing of its properties in dependence on the choice of $\mathcal{C}=e^{\chi{i}\mathcal{R}\mathcal{P}_\xi}\mathcal{P}_\xi$ in (\ref{bonn22}).

Our first result in this direction deals with the description of operators $T$
which determine (with the help of (\ref{eee8})) self-adjoint operators in the Krein space $(\mathfrak{H}, [\cdot,\cdot]_{\mathcal{P}_\xi})$
\emph{with nonnegative real spectrum}.
\begin{proposition}\label{bonn70b}
The following statements are equivalent:
\begin{itemize}
  \item[(i)] a self-adjoint extension $H$ of $B^2$ in the Krein space $(\mathfrak{H}, [\cdot,\cdot]_{\mathcal{P}_\xi})$
has a nonnegative real spectrum and $H\mathcal{C}=\mathcal{C}H$ on $\mathcal{D}(H)$  for a certain choice of  $\mathcal{C}=e^{\chi{i}\mathcal{R}\mathcal{P}_\xi}\mathcal{P}_\xi$;
  \item[(ii)] the operator $H$ is defined by the formula
\begin{equation}\label{eee8b}
  H={B^*}^2\upharpoonright_{\mathcal{D}(H)}, \qquad  \mathcal{D}(H)=\{f\in{\mathcal{D}({B^{*}}^2)} \ :  \  T\Gamma_{1}f=\Gamma_{0}f\},
    \end{equation}
where $T$ is a bounded self-adjoint operator in the Krein space $(\mathcal{H}, [\cdot,\cdot]_{\mathcal{P}_\xi})$, which commutes with the
operator $\mathcal{C}=e^{\chi{i}\mathcal{R}\mathcal{P}_\xi}\mathcal{P}_\xi$ and satisfies the inequality
\begin{equation}\label{bonn193}
0\leq{e^{-\chi{i}\mathcal{R}\mathcal{P}_\xi}}T\leq\frac{1}{2}{e^{-\chi{i}\mathcal{R}\mathcal{P}_\xi}},
\end{equation}
which is understood with respect to $(\cdot,\cdot)$.
\end{itemize}
\end{proposition}
\emph{Proof.} Remembering (\ref{eee8}), we conclude that the formula (\ref{eee8b}) determines a collection of intermediate extensions
$H$ of $B^2$ with $-1\in\rho(H)$ when the parameter $T$ runs the set of bounded operators in $\mathcal{H}$.
 According to Lemma \ref{es16}, a subset of self-adjoint extensions $H$ of $B^2$ in the Krein space $(\mathfrak{H}, [\cdot,\cdot]_{\mathcal{P}_\xi})$ is distinguished by the additional condition that  $T$ is a self-adjoint operator in the Krein space $(\mathcal{H}, [\cdot,\cdot]_{\mathcal{P}_\xi})$.

Since the boundary operators $\Gamma_j$ satisfy (\ref{bonn44b}) and (\ref{es45}),
the operator $\mathcal{C}=e^{\chi{i}\mathcal{R}\mathcal{P}_\xi}\mathcal{P}_\xi$ commutes with $\Gamma_j$.
Repeating the proof of Lemma \ref{bonn43} with the use of $\mathcal{C}$ instead of $\mathcal{PT}$,
we deduce that the relation $\mathcal{C}Hf=H\mathcal{C}f$ holds for all $f\in\mathcal{D}(H)$ if and only if the identity  $\mathcal{C}T=T\mathcal{C}$ holds on $\mathcal{H}$.
Therefore, the operators $H$ and $T$ turn out to be self-adjoint, respectively, in the Hilbert spaces $\mathfrak{H}$ and $\mathcal{H}$
endowed with the inner product $(\cdot,\cdot)_{\mathcal{C}}$. Furthermore, $H$ is a nonnegative self-adjoint extension of $B^2$,
due to the assumption about nonnegativity of the spectrum of $H$.

Using the well-known result \cite{Krein} on extremal properties of the Friedrichs $H_\mu=B^*B$ and
the Krein -- von Neumann $H_M=BB^*$ extensions of $B^2$, we arrive at the conclusion that the self-adjoint extension $H$ of $B^2$
 is nonnegative if and only if
\begin{equation}\label{bonn90}
(H_\mu+I)^{-1}\le{(H+I)^{-1}}\leq{(H_M+I)^{-1}},
\end{equation}
where the operator inequalities are understood with respect to $(\cdot,\cdot)_{\mathcal{C}}$.

Since $T=[({H}+{I})^{-1}-(H_{\mu}+{I})^{-1}]\upharpoonright_{\mathcal H}$, the inequalities (\ref{bonn90}) can be
rewritten in the following equivalent form
$$
0\le{T}\leq{T_M}=(H_M+I)^{-1}-(H_{\mu}+I)^{-1},
$$
where, without loss of generality, we assume that $Tf=T_Mf=0$ for all elements $f\in\mathfrak{H}\ominus\mathcal{H}=\mathcal{R}(B^2+I)$.

It is easy to check (using the presentation (\ref{e14}) and the explicit formulas (\ref{bonn41}), (\ref{eee7}) for $\Gamma_j$) that
$T_M=\frac{1}{2}I$. Therefore, the operator $T$ defines a nonnegative self-adjoint extension $H$ of $B^2$ in (\ref{eee8b})
if and only if $0\le{T}\leq\frac{1}{2}I$ with respect to $(\cdot,\cdot)_{\mathcal{C}}$. Using the explicit formula
(\ref{bonn22}) for $(\cdot,\cdot)_{\mathcal{C}}$ we rewrite the obtained inequality in the form (\ref{bonn193}), which corresponds
to the case of initial inner product $(\cdot,\cdot)$.
Proposition \ref{bonn70b} is proved. \rule{2mm}{2mm}
\begin{corollary}\label{esse89}
Let the deficiency indices of $B^2$ be equal $<2,2>$ and let $H$ be defined by (\ref{eee8b}) with
$T=\beta_0{I}+\beta_1\mathcal{C}$, where
$\mathcal{C}=e^{\chi{i}\mathcal{R}\mathcal{P}_\xi}\mathcal{P}_\xi$  and $\beta_0, \ \beta_1\in\mathbb{R}$.
Then $H$ has nonnegative real spectrum if and only if
\begin{equation}\label{bonn305}
0\leq\beta_0\leq\frac{1}{2}  \quad \mbox{and} \quad |\beta_1|\leq\textsf{min}\{\frac{1}{2}-\beta_0, \beta_0\}
\end{equation}
\end{corollary}

\emph{Proof.} It follows from Lemma \ref{es16}, and Corollary \ref{esse8} that the operator $H$ defined by
(\ref{eee8b}) with $T=\beta_0{I}+\beta_1\mathcal{C}$ is a self-adjoint extension of $B^2$ in the Krein space $(\mathfrak{H}, [\cdot,\cdot]_{\mathcal{P}_\xi})$ and such that the identity $H\mathcal{C}=\mathcal{C}H$ holds on $\mathcal{D}(H)$.

By virtue of Proposition \ref{bonn70b}, the operator $H$ has a nonnegative real spectrum if and only if inequality
(\ref{bonn193}) holds, that is
\begin{equation}\label{bonn400}
0\leq{e^{-\chi{i}\mathcal{R}\mathcal{P}_\xi}}[\beta_0{I}+\beta_1\mathcal{C}]={\beta_0}e^{-\chi{i}\mathcal{R}\mathcal{P}_\xi}+\beta_1\mathcal{P}_\xi\leq\frac{1}{2}{e^{-\chi{i}\mathcal{R}\mathcal{P}_\xi}}
\end{equation}

Since the dimension of $\mathcal{H}=\ker({B^*}^2+I)$ is $2$ and the restrictions of $\mathcal{P}_\xi$ and $\mathcal{R}$ onto $\mathcal{H}$ are anti-commuting involutions in $\mathcal{H}$, there exists an orthonormal basis $h_0, h_1$ of $\mathcal{H}$ such that $\mathcal{P}_{\xi}h_j=(-1)^jh_j$ and $\mathcal{R}h_0=h_1$. This means that the operators $\mathcal{P}_\xi$ and $\mathcal{R}$ in $\mathcal{H}$
can be identified with the Pauli matrices $\sigma_3=\left(\begin{array}{cc}
1 & 0 \\
0 & -1 \end{array}\right)$ and $\sigma_1=\left(\begin{array}{cc}
0 & 1 \\
1 & 0 \end{array}\right)$, respectively. Then $i\mathcal{RP}_\xi$ corresponds to $\sigma_2=\left(\begin{array}{cc}
0 & -i \\
i & 0 \end{array}\right)$ and
$$
{\beta_0}e^{-\chi{i}\mathcal{R}\mathcal{P}_\xi}+\beta_1\mathcal{P}_\xi\ {\leftrightarrow}\
\left(\begin{array}{cc}
\beta_0\cosh\chi+\beta_1 & i\beta_0\sinh\chi \\
-i\beta_0\sinh\chi & \beta_0\cosh\chi-\beta_1
\end{array}\right).
$$

The obtained matrix corresponds to a nonnegative operator if and only if the equation
$$
\lambda^2-2\lambda\beta_0\cosh\chi+\beta_0^2-\beta_1^2=0
$$
has nonnegative roots. This takes place if $\beta_0\geq{0}$ and $|\beta_1|\leq\beta_0$.
In that case, the self-adjoint operator ${\beta_0}e^{-\chi{i}\mathcal{R}\mathcal{P}_\xi}+\beta_1\mathcal{P}_\xi$
is nonnegative in $\mathcal{H}$.

Rewriting ${\beta_0}e^{-\chi{i}\mathcal{R}\mathcal{P}_\xi}+\beta_1\mathcal{P}_\xi\leq\frac{1}{2}{e^{-\chi{i}\mathcal{R}\mathcal{P}_\xi}}$
as $0\leq{\beta_0'}e^{-\chi{i}\mathcal{R}\mathcal{P}_\xi}+\beta_1'\mathcal{P}_\xi$, where
$\beta_0'=\frac{1}{2}-\beta_0$, $\beta_1'=-\beta_1$ and repeating the previous arguments we obtain
$\frac{1}{2}-\beta_0\geq{0}$ and $|\beta_1|\leq{\frac{1}{2}-\beta_0}$. Therefore, the relations (\ref{bonn305}) and (\ref{bonn400})
are equivalent. Corollary \ref{esse89} is proved.
\rule{2mm}{2mm}

\subsection{Properties of scattering matrix.}
An arbitrary operator $H$ satisfying conditions (i) of Proposition \ref{bonn70b} turns out to be $0$-perturbed in the sense of Definition \ref{ddd6}
(with respect to $(\cdot,\cdot)_{\mathcal{C}}$).
\emph{For such a kind of operators, the scattering matrix $\mathbb{S}(\delta)$ and its analytical continuation $\mathbb{S}(z)$ in the lower half-plane
are defined by formulas (\ref{bonn65b}) and (\ref{bonn65}), respectively}. These functions possess additional properties due to
the properties of operators $T$ from item (ii) of Proposition \ref{bonn70b}.
Precisely:

{\bf 1.} Since $T$ is a bounded self-adjoint operator in $(\mathcal{H}, [\cdot,\cdot]_{\mathcal{P}_\xi})$, the identity
${\mathcal{P}_\xi}T=T^*{\mathcal{P}_\xi}$ holds. Combining this relation with (\ref{bonn65b}) we obtain that
\begin{equation}\label{bonn72}
\mathcal{P}_\xi\mathbb{S}(\delta)=\mathbb{S}^*(-\delta)\mathcal{P}_\xi.
\end{equation}

{\bf 2.} Since $T$ commutes with $\mathcal{C}$ the identity
 ${\mathcal{C}}\mathbb{S}(\delta)=\mathbb{S}(\delta){\mathcal{C}}$ holds. Combining it with (\ref{bonn72}) and taking into account that
 $\mathcal{C}=e^{\chi{i}\mathcal{R}\mathcal{P}_\xi}\mathcal{P}_\xi$ we obtain that
\begin{equation}\label{bonn73}
e^{-\chi{i}\mathcal{R}\mathcal{P}_\xi}\mathbb{S}(\delta)=\mathbb{S}^*(-\delta)e^{-\chi{i}\mathcal{R}\mathcal{P}_\xi}.
\end{equation}

For the analytical continuation $\mathbb{S}(z)$, the relations (\ref{bonn72}) and (\ref{bonn73}) are transformed as follows:
$$
\mathcal{P}_\xi\mathbb{S}(z)=\mathbb{S}^*(-\overline{z})\mathcal{P}_\xi, \qquad e^{-\chi{i}\mathcal{R}\mathcal{P}_\xi}\mathbb{S}(z)=\mathbb{S}^*(-\overline{z})e^{-\chi{i}\mathcal{R}\mathcal{P}_\xi}.
$$

The scattering matrix $\mathbb{S}(\delta)$ admits an `external' description in the sense of the following Theorem.
\begin{theorem}\label{tetete3}
Let $B$ be a simple maximal symmetric operator in $\mathfrak{H}$ which commutes with the elements of the Clifford algebra
$\mathcal{C}l_2(\mathcal{P}, \mathcal{R})$. Then
an operator-valued function ${\mathbb S}(\delta)$, the values of
which are bounded operators in a Hilbert space $\mathcal{H}=\ker({B^*}^2+I)$, is the scattering matrix for some choice of an operator $H$
satisfying the condition of item (i) of Proposition \ref{bonn70b} if and only if the following conditions
hold:

(a) \ the function ${\mathbb S}(\delta)$ is the boundary value in the sense of
strong convergence of an operator-valued function ${\mathbb S}(z)$ analytic in
the lower half-plane and satisfying the inequality
$$
{\mathbb S}^*(z)e^{-\chi{i}\mathcal{R}\mathcal{P}_\xi}{\mathbb S}(z){\leq}e^{-\chi{i}\mathcal{R}\mathcal{P}_\xi}, \qquad \forall{z}\in\mathbb{C}_-;
$$

(b) \ the identity
$e^{-\chi{i}\mathcal{R}\mathcal{P}_\xi}\mathbb{S}(z)=\mathbb{S}^*(-\overline{z})e^{-\chi{i}\mathcal{R}\mathcal{P}_\xi}$ holds for all
$z\in\mathbb{C}_-$;

(c) \ the identity
$$
(Re \ z)[e^{-\chi{i}\mathcal{R}\mathcal{P}_\xi}-{\mathbb S}^*(z)e^{-\chi{i}\mathcal{R}\mathcal{P}_\xi}{\mathbb S}(z)]=i(Im \ z)[{\mathbb S}^*(z)e^{-\chi{i}\mathcal{R}\mathcal{P}_\xi}-e^{-\chi{i}\mathcal{R}\mathcal{P}_\xi}{\mathbb S}(z)]
$$
is true for at least one $z\in\mathbb{C}_-$ with $Re \ z\not=0$;

(d) \ the identity $\mathcal{P}_\xi\mathbb{S}(z)=\mathbb{S}^*(-\overline{z})\mathcal{P}_\xi$ holds for at least one $z\in\mathbb{C}_-$.
\end{theorem}
\emph{Proof.}
 Let $H$ satisfy the condition of item (i) of Proposition \ref{bonn70b}. Then $H$  is
a nonnegative self-adjoint extension of $B^2$ (that is, $0$-perturbed operator)
 in the Hilbert space $\mathfrak{H}$ endowed with the inner product $(\cdot,\cdot)_{\mathcal{C}}$.
In that case, the scattering matrix corresponding to $H$ has an analytical continuation ${\mathbb S}(z)$ in the lower half-plane
and characteristic properties of ${\mathbb S}(z)$ are known \cite[Theorems 4.1, 4.2]{AlAn}.
We formulate them using the notation ${\mathbb S}^{[*]}(\cdot)$ for the adjoint of ${\mathbb S}(\cdot)$ with respect to the inner
product $(\cdot,\cdot)_{\mathcal{C}}$ in $\mathcal{H}$. Namely:
$$
(a') \quad  {\mathbb S}^{[*]}(z){\mathbb S}(z)\leq{I}; \qquad
(b') \quad  \mathbb{S}(z)={\mathbb S}^{[*]}(-\overline{z})
$$
for all $z\in\mathbb{C}_-$ and
$$
(c') \quad (Re \ z)[I-{\mathbb S}^{[*]}(z){\mathbb S}(z)]=i(Im \ z)[{\mathbb S}^{[*]}(z)-{\mathbb S}(z)]
$$
for at least one $z\in\mathbb{C}_-$ with $Re \ z\not=0$.

It follows from (\ref{bonn22}) that
\begin{equation}\label{bonn95}
{\mathbb S}^{[*]}(\cdot)=e^{\chi{i}\mathcal{R}\mathcal{P}_\xi}{\mathbb S}^*(\cdot)e^{-\chi{i}\mathcal{R}\mathcal{P}_\xi},
\end{equation}
where ${\mathbb S}^*(\cdot)$ is the adjoint of ${\mathbb S}(\cdot)$ with respect to the initial inner product $(\cdot,\cdot)$ in $\mathcal{H}$.
By virtue of (\ref{bonn95}) the equivalence between $(a), (b), (c)$ and $(a'), (b'), (c')$ is obvious.

Since $H$ satisfies the condition of item (i) of Proposition \ref{bonn70b}, the bounded operator $T$ in (\ref{eee8b})
is self-adjoint in the Krein space $(\mathcal{H}, [\cdot,\cdot]_{\mathcal{P}_\xi})$, that is the relation $T{\mathcal{P}_\xi}={\mathcal{P}_\xi}T^*$ holds on $\mathcal{H}$.
This relation and the explicit formula (\ref{bonn65}) for  ${\mathbb S}(z)$ imply $(d)$.
Thus, if $H$ satisfies item (i) of Proposition \ref{bonn70b}, then the analytical continuation of the scattering matrix
${\mathbb S}(\cdot)$ satisfies conditions $(a)-(d)$.

Conversely, assume that an operator-function ${\mathbb S}(\delta)$ has an analytical continuation ${\mathbb S}(z)$ in $\mathbb{C}_-$, which satisfies
$(a)-(d)$.  The conditions $(a)-(c)$ are equivalent to the conditions $(a')-(c')$, which mean that  ${\mathbb S}(\delta)$ is the scattering
matrix for a nonnegative self-adjoint extension $H$ of $B^2$ in the Hilbert space $\mathfrak{H}$ \emph{endowed with the inner product} $(e^{-\chi{i}\mathcal{R}\mathcal{P}_\xi}\cdot, \cdot)$; \cite[Theorems 4.1, 4.2]{AlAn}. In this case, ${\mathbb S}(z)$ is determined by (\ref{bonn65}), where $T$ is a bounded operator in $\mathcal{H}$. The operator $T$ defines $H$ via formula (\ref{eee8b}) and
it can be expressed (from (\ref{bonn65})) as
\begin{equation}\label{bonn99}
T=\frac{1}{2(iz-1)}(I-{\mathbb S}(z))({\mathbb S}(z)-\theta(z){I})^{-1}, \qquad \theta(z)=\frac{1+iz}{1-iz},
\end{equation}
where the right-hand side does not depend on $z\in\mathbb{C}_-$ (see \cite{AlAn} for details).

Using condition $(d)$, we deduce from (\ref{bonn99}) that $T$ satisfies the relation ${\mathcal{P}_\xi}T=T^*{\mathcal{P}_\xi}$ on $\mathcal{H}$. Therefore, $T$ is self-adjoint in the Krein space $(\mathcal{H}, [\cdot,\cdot]_{\mathcal{P}_\xi})$.
Then, by Lemma \ref{es16}, $H$ is a self-adjoint operator in the Krein space $(\mathfrak{H}, [\cdot,\cdot]_{\mathcal{P}_\xi})$.
Thus, ${\mathcal{P}_\xi}H=H^*{\mathcal{P}_\xi}$ on $\mathcal{D}(H)$.

On the other hand, as was mentioned above, $H$ is a self-adjoint operator with respect to the inner product $(e^{-\chi{i}\mathcal{R}\mathcal{P}_\xi}\cdot, \cdot)$ in $\mathfrak{H}$.
This means that the relation
$e^{-\chi{i}\mathcal{R}\mathcal{P}_\xi}H=H^*e^{-\chi{i}\mathcal{R}\mathcal{P}_\xi}$ holds on $\mathcal{D}(H)$,
where $H^*$ is the adjoint of $H$ with respect to the initial inner product $(\cdot,\cdot)$.
Here $e^{-\chi{i}\mathcal{R}\mathcal{P}_\xi}$ maps $\mathcal{D}(H)$ onto $\mathcal{D}(H^*)$ and, hence, the latter relation is
equivalent to the relation $e^{\chi{i}\mathcal{R}\mathcal{P}_\xi}H^*=He^{\chi{i}\mathcal{R}\mathcal{P}_\xi}$, which holds on $\mathcal{D}(H^*)$.

Combining the obtained operator identities we conclude that
$$
{\mathcal C}Hf=e^{\chi{i}\mathcal{R}\mathcal{P}_\xi}\mathcal{P}_\xi{H}f=e^{\chi{i}\mathcal{R}\mathcal{P}_\xi}H^*\mathcal{P}_\xi{f}=He^{\chi{i}\mathcal{R}\mathcal{P}_\xi}\mathcal{P}_\xi{f}=H{\mathcal C}f
$$
for all elements $f\in\mathcal{D}(H)$. Thus $H$ commutes with ${\mathcal C}$. Summing up the properties of $H$ established above we conclude that $H$ satisfies the condition of item (i)
of Proposition \ref{bonn70b}.  Theorem \ref{tetete3} is proved. \rule{2mm}{2mm}

\begin{corollary}\label{tetete4}
 Let $H$  be an operator satisfying item (i) of Proposition \ref{bonn70b}.  Then $H$ is $\mathcal{PT}$-symmetric
if and only if the analytical continuation $\mathbb{S}(z)$ of its scattering matrix satisfies the relation
\begin{equation}\label{bonn97}
\mathcal{PT}\mathbb{S}(z)=\mathbb{S}(-\overline{z})\mathcal{PT}, \qquad \forall{z}\in\mathbb{C}_-.
\end{equation}
\end{corollary}
\emph{Proof.} If $H$ satisfies item (i) of Proposition \ref{bonn70b}, then its scattering matrix has an analytical continuation $\mathbb{S}(z)$ with properties $(a)-(d)$. Furthermore, ${\mathbb S}(z)$ is determined by (\ref{bonn65}) and formula (\ref{bonn99}) holds.

At the beginning of Section 4 we suppose that the operator $B$ possesses properties (\ref{zvit1}). Then the operators
${B^*}^2$ and $H_\mu$ commute with $\mathcal{PT}$ due to (\ref{zvit3}). This means that the boundary operators $\Gamma_j$ in (\ref{eee7})
commute with $\mathcal{PT}$. Therefore, $H$ is $\mathcal{PT}$-symmetric in $\mathfrak{H}$ if and only if the corresponding operator $T$ is
$\mathcal{PT}$-symmetric in $\mathcal{H}$; Lemma \ref{bonn43}. In this case, formula (\ref{bonn65}) shows that $\mathbb{S}(z)$ satisfies
(\ref{bonn97}).

Conversely, if (\ref{bonn97}) holds, then $\mathcal{PT}T=T\mathcal{PT}$ due to (\ref{bonn99}) and $H$ is $\mathcal{PT}$-symmetric thanks to
Lemma \ref{bonn43}. Corollary \ref{tetete4} is proved. \rule{2mm}{2mm}
\subsection{One dimensional Schr\"{o}dinger operator with $\mathcal{PT}$-symmetric zero-range potential.}
Let $\mathcal{P}$ be the space parity operator: $\mathcal{P}f(x)=f(-x)$ and let $\mathcal{T}$ be the complex conjugation: $\mathcal{T}f(x)=\overline{f(x)}$ in $\mathfrak{H}=L_2(\mathbb{R})$.

The unitary involutions $\mathcal{P}$ and $\mathcal{R}=(\textsf{sgn}\ x)I$ anti-commute in $L_2(\mathbb{R})$ and these operators are generators
of the Clifford algebra $\mathcal{C}l_2(\mathcal{P}, \mathcal{R})$.

The operator
$$
B=(\textsf{sgn}\ x)i\frac{d}{dx}, \qquad \mathcal{D}(B)=\{ u\in{{W}_{2}^{1}}({\mathbb R}\setminus\{0\}) \ : \ u(\pm{0})=0 \}.
$$
is simple maximal symmetric in $L_2(\mathbb{R})$ and it satisfies (\ref{zvit1}).
Then the symmetric operator
$$
B^2=-\frac{d^2}{dx^2}, \qquad \mathcal{D}(B^2)=\{ u\in{W}_{2}^{2}({\mathbb R}\setminus\{0\}) \ : \ u(\pm{0})=u'(\pm{0})=0 \}
$$
and its adjoint
$$
{B^2}^*={B^*}^2=-\frac{d^2}{dx^2}, \qquad \mathcal{D}({B^*}^2)={W}_{2}^{2}({\mathbb R}\setminus\{0\}).
$$
satisfy (\ref{zvit3}).

Intermediate\footnote{that is, extensions of $B^2$ and, simultaneously, restrictions of ${B^*}^2$} $\mathcal{PT}$-symmetric extensions of $B^2$  can be interpreted as one-dimensional Schr\"{o}dinger operators corresponding to the
formal expression $-\frac{d^2}{dx^2}+V(x)$ with a general zero-range $\mathcal{PT}$-symmetric potential $V(x)$ concentrated at the point $x=0$
\cite{AlKur, AlKuz}.

The Friedrichs extension $H_{\mu}=B^*B$ of $B^2$ has the form
$$
H_\mu=-\frac{d^2}{dx^2}, \qquad \mathcal{D}(H_\mu)=\{u\in{W}_{2}^{2}({\mathbb R}\setminus\{0\}) \ : \ u(\pm{0})=0 \}
$$
and the boundary operators $\Gamma_j : \mathcal{D}({B^*}^2)\to\mathcal{H}$ defined by (\ref{eee8}) act as follows
$$
\begin{array}{l}
\Gamma_0f(x)=\frac{1}{2}[f(+0)+f(-0)]h_1+\frac{1}{2}[f(+0)-f(-0)]h_2 \vspace{3mm} \\
\Gamma_1f(x)=2\Gamma_0f(x)+[f'(+0)-f'(-0)]h_1+[f'(+0)+f'(-0)]h_2,
\end{array}
$$
where the functions
$$
h_1(x)=\left\{\begin{array}{cc}
 e^{{-x}}, & x>0  \\
 e^{x}, & x<0
 \end{array}\right.    \qquad
 h_{2}(x)=\left\{\begin{array}{cc}
 e^{-x}, & x>0  \\
 -e^{x}, & x<0
 \end{array}\right.
$$
form an orthogonal basis of $\mathcal{H}=\ker({B^*}^2+I)$.

Identifying $h_1$ and $h_2$ with $\left(\begin{array}{c}
1 \\
0 \end{array}\right)$ and with $\left(\begin{array}{c}
0 \\
1 \end{array}\right)$, respectively, we can identify the operators $\Gamma_j$ above with
the operators
\begin{equation}\label{bonn100}
\Gamma_0f(x)=\frac{1}{2}\left(\begin{array}{c}
f(+0)+f(-0) \\
f(+0)-f(-0)
\end{array}\right), \ \Gamma_1f(x)=2\Gamma_0f(x)+\left(\begin{array}{c}
f'(+0)-f'(-0) \\
f'(+0)+f'(-0)
\end{array}\right)
\end{equation}
mapping $\mathcal{D}({B^*}^2)$ onto $\mathbb{C}^2$ (we preserve the same notation for $\Gamma_j$).

Under such the identification of $\mathcal{H}$ with $\mathbb{C}^2$, the restriction of $\mathcal{P}$ and $\mathcal{R}$ onto
$\mathcal{H}$ coincide with the Pauli matrices $\sigma_3=\left(\begin{array}{cc}
1 & 0 \\
0 & -1 \end{array}\right)$ and $\sigma_1=\left(\begin{array}{cc}
0 & 1 \\
1 & 0 \end{array}\right)$, respectively. Then $i\mathcal{RP}$ corresponds to $\sigma_2=\left(\begin{array}{cc}
0 & -i \\
i & 0 \end{array}\right)$ and
\begin{equation}\label{bonn504}
\mathcal{P}_\xi=e^{i\xi\mathcal{R}}\mathcal{P}\ {\leftrightarrow}\ \sigma_{3\xi}=e^{i\xi\sigma_1}\sigma_3=\left(\begin{array}{cc}
\cos\xi & -i\sin\xi \\
i\sin\xi & -\cos\xi
\end{array}\right).
\end{equation}

It follows from Theorem \ref{esse8} and the identification above that
the formula (\ref{eee8}) with boundary mappings $\Gamma_j$ defined by (\ref{bonn100})
and with matrices
\begin{equation}\label{bonn500}
T=\beta_0\sigma_0+\beta_1e^{\chi{i}\sigma_1\sigma_{3\xi}}\sigma_{3\xi}, \quad \sigma_0=\left(\begin{array}{cc}
1 & 0 \\
0 & 1 \end{array}\right), \quad \beta_0, \beta_1\in\mathbb{R}
\end{equation}
determines $\mathcal{PT}$-symmetric extensions $H$ of $B^2$, which are self-adjoint in the Krein space
$(\mathfrak{H}, [\cdot,\cdot]_{\mathcal{P}_\xi})$ and they have the property of $\mathcal{C}$-symmetry with  $\mathcal{C}=e^{\chi{i}\mathcal{R}\mathcal{P}_\xi}\mathcal{P}_\xi$.
Therefore, these operators are self-adjoint in the Hilbert space $L_2(\mathbb{R})$ with the new inner product
(\ref{bonn22}).

The trace $\textsf{tr}(T)$ and the determinant $\det(T)$ of the matrix $T$ in (\ref{bonn500}) are related with the parameters
$\beta_0, \beta_1$ as follows:
$$
\beta_0=\frac{1}{2}\textsf{tr}(T), \qquad \beta_0^2-\beta_1^2=\det(T).
$$

The self-adjointness of $H$ ensures the reality of the corresponding spectra $\sigma(H)$.
Employing Corollary \ref{esse89} we conclude that $\sigma(H)$ is nonnegative if parameters $\beta_0, \beta_1$
in (\ref{bonn500}) satisfy (\ref{bonn305}). In that case, the operators $H$ turn out to be $0$-perturbed in the sense of Definition \ref{ddd6}
(with respect to the new inner product (\ref{bonn22})) and the analytic continuations of the corresponding scattering matrices (\ref{bonn65})
take the form
\begin{equation}\label{bonn503}
\mathbb{S}(z)=\frac{[1-2(1+iz)\beta_0]\sigma_{3\xi}-[2(1+iz)\beta_1]e^{\chi{i}\sigma_1\sigma_{3\xi}}}{[1-2(1-iz)\beta_0]\sigma_{3\xi}-[2(1-iz)\beta_1]e^{\chi{i}\sigma_1\sigma_{3\xi}}}, \qquad z\in\mathbb{C}_-.
\end{equation}

In (\ref{bonn503}), the matrix $\sigma_{3\xi}$ contains an information about the Krein space $(L_2(\mathbb{R}), [\cdot,\cdot]_{\mathcal{P}_\xi})$ in which a
$\mathcal{PT}$-symmetric operator $H$ can be interpreted as a self-adjoint one (thanks to (\ref{bonn504})), while $e^{\chi{i}\sigma_1\sigma_{3\xi}}$ characterizes
the new inner product (\ref{bonn22}) of $L_2(\mathbb{R})$ with respect to which $H$ turns out to be self-adjoint
as an operator acting in a Hilbert space.

The values of $\mathbb{S}(z)$ are contraction operators in $\mathbb{C}^2$ if and only if $\beta_1=0$.
In that case, the operators $H$ determined by (\ref{eee8}) with $T=\beta_0\sigma_0$, $0\leq\beta_0\leq\frac{1}{2}$
are self-adjoint extensions of $B^2$ which have nonnegative real spectra and commute with every element of the Clifford
algebra $\mathcal{C}l_2(\mathcal{P}, \mathcal{R})$.
The Friedrichs and the Krein-von Neumann extensions of $B^2$ are distinguished by the endpoints $\beta_0=0$ and $\beta_1=\frac{1}{2}$, respectively, and the corresponding functions $\mathbb{S}(z)$ coincide with $\sigma_0$ and $-\sigma_0$.

\smallskip

\noindent \textbf{Acknowledgements.}
The financial support by the DFG-project AL 214/33-1 (both authors) and JRP IZ73Z0 (28135) of SCOPES 2009-2012 (the second named author)
is gratefully acknowledged.

\end{document}